\documentclass[particles,review,submit,pdftex,moreauthors]{Definitions/mdpi} 

\firstpage{1} 
\pubvolume{1}
\issuenum{1}
\articlenumber{0}
\pubyear{2026}
\copyrightyear{2026}
\datereceived{ } 
\daterevised{ } % Comment out if no revised date
\dateaccepted{ } 
\datepublished{ }

\Title{Recent developments and  applications of the relativistic chiral nuclear force}

\Author{Li-Sheng Geng $^{1,2,3,4}$\orcidA{}, Jun-Xu Lu$^{1}$, Qing-Yu Zhai$^{1}$, Zhi-Wei Liu$^{1}$, and Shihang Shen$^{3}$\orcidE{}}

\AuthorNames{Li-Sheng Geng, Jun-Xu Lu, Qing-Yu Zhai, Zhi-Wei Liu, and Shihang Shen}

\address{%

$^{1}$\quad School of Physics, Beihang University, Beijing 102206, China\\

$^{2}$\quad Sino-French Carbon Neutrality Research Center, \'Ecole Centrale de P\'ekin/School of General Engineering, Beihang University, Beijing 100191, China\\

$^{3}$\quad Peng Huanwu Collaborative Center for Research and Education, Beihang University, Beijing 100191, China\\

$^{4}$\quad  Southern Center for Nuclear-Science Theory (SCNT), Institute of Modern Physics, Chinese Academy of Sciences, Huizhou 516000, China
}
\corres{Correspondence: lisheng.geng@buaa.edu.cn; Tel.:  +86-10-6171-6750 }

\abstract{
The nuclear force is central to our understanding of complex nuclear phenomena and to the applications of nuclear techniques. The nonperturbative nature of the low-energy strong interaction and the color confinement have made an ab initio understanding of the nuclear force a challenge for almost a century since the pioneering work of Yukawa. Since 1990, chiral effective field theory (ChEFT) has become the de facto standard for describing nuclear interactions—most prior studies employed heavy-baryon chiral perturbation theory. Only recently, there have been successful attempts to construct a chiral nuclear force employing covariant baryon chiral perturbation theory. In this work, we review recent developments and applications of relativistic chiral nuclear forces. We first elaborate on the necessity of relativistic/covariant theories, then present the construction of the first high-precision relativistic chiral nuclear force up to next-to-next-to-leading order (NNLO), and discuss the ongoing progress in higher-order nucleon-nucleon (NN) and n-$d$ scattering, as well as their applications in nuclear matter, finite nuclei, and hypernuclear systems. Finally, we summarize the achievements and outline the future outlook of this research field.}


\begin{document}

%%%%%%%%%%%%%%%%%%%%%%%%%%%%%%%%%%%%%%%%%%

\section{Introduction}
The nuclear force is fundamental to understanding the rich and intricate phenomena of nuclear physics, nuclear astrophysics, exotic hadron states, and new physics beyond the standard model.  Because of the non-perturbative nature of the strong interaction and color confinement, understanding the nuclear force has remained a central topic in the nuclear physics and strong-interaction communities. For instance, in 2020, to celebrate its 50th anniversary, Physical Review C selected a collection of milestone papers that remain central to nuclear physics. These papers announce major discoveries or open up new avenues of research (\url{https://journals.aps.org/prc/50th}). Seven of the 41 selected are dedicated to developing high-precision nuclear forces. In addition, many rare isotope beam facilities around the world, such as RIBF (RI Beam Factory)~\cite{Yano:2007zz}, HIAF (High Intensity heavy-ion Accelerator Facility)~\cite{Yang:2013yeb}, and  FRIB (Facility for Rare Isotope Beams)~\cite{Gade:2016xrp}, list studying the nuclear force as one of their major scientific goals. 

Attempts to understand the nuclear force dated back to the 1930s, when Yukawa proposed the pion-exchange theory~\cite{Yukawa:1935xg}. Later in the 1950s and 1960s, with the experimental discovery of the pion and other heavier mesons (such as $\rho$ and $\omega$), the one-boson exchange model was developed~\cite{DeTourreil:1975gz, Holinde:1976mkn, Nagels:1977ze}, which qualitatively described the properties of the nuclear force and the deuteron. In the 1970s, Quantum Chromodynamics (QCD) was established as the fundamental theory of the strong interaction~\cite{Gross:1973id,Politzer:1973fx}, whose basic degrees of freedom are quarks and gluons. Then, physicists attempted to explain the nuclear force at the quark level. Various types of quark models~\cite{Oka:1980ax,Faessler:1982ik,Shimizu:1989ye, Valcarce:2005em,Straub:1988gj,Zhang:1994pp,Zhang:1997ny,Dai:2003dz,Wang:1992wi,Wu:1996fm,Ping:1998si,Wu:1998wu,Pang:2001xx,Fujiwara:1995td,Fujiwara:1995fx,Fujiwara:1996qj,Fujita:1998sg} were proposed. However, these models, based on quark degrees of freedom, are, in principle, phenomenological, and their connection to QCD remains unclear.

Beginning in 1990, Weinberg proposed that the chiral effective field theory (ChEFT) derived from QCD could be used to describe the nuclear force~\cite{Weinberg:1990rz, Weinberg:1991um, Weinberg:1992yk}. Unlike the phenomenological nuclear forces (such as AV18~\cite{Wiringa:1994wb}, Reid93~\cite{Stoks:1994wp}, and CD-Bonn~\cite{Machleidt:2000ge}), Weinberg's chiral nuclear force is intimately related to the fundamental theory of the strong interaction, QCD, making it a microscopic theory. In the chiral effective field theory, through the power counting, the relative magnitude of a certain part of the nuclear force can be estimated a priori, thereby achieving a systematic order-by-order calculation, which provides a method to systematically improve the description according to the chiral orders and estimate the theoretical uncertainties~\cite{Epelbaum:2008ga,Meissner:2014lgi,Hammer:2019poc}. Besides, the power counting of the chiral effective field theory can consistently incorporate three-body and even four-body interactions, which naturally emerge since the nucleons are not point-like elementary particles but composite particles with internal structures~\cite{Hebeler:2020ocj}. For the early history and literature on the development of nuclear forces, we refer to the comprehensive reviews ~\cite{Epelbaum:2008ga,Machleidt:2011zz,Machleidt:2024bwl}. 

After more than 30 years of development, the nonrelativistic chiral nuclear force has proven to be highly successful. It has become the de facto standard input for \textit{ab initio}  nuclear studies, thereby pioneering the study of non-perturbative strong interactions within effective field theories. Weinberg constructed the leading-order (LO) chiral nuclear force, which includes the long-range one-pion-exchange (OPE) term (Yukawa interaction) and short-range contact terms~\cite{Weinberg:1990rz,Weinberg:1991um}. Later, van Kolck et al. extended it to next-to-next-to-leading order (NNLO) via time-ordered chiral effective field theory, adding two-pion exchanges (TPE), relativistic corrections, and new contact terms~\cite{Ordonez:1993tn,Ordonez:1995rz}.
Epelbaum et al. noted the energy dependence of this force and proposed a new NNLO scheme using unitary transformations~\cite{Epelbaum:1998ka,Epelbaum:1999dj}, later improving its convergence with spectrum function regularization~\cite{Epelbaum:2003gr,Epelbaum:2003xx}. 
Machleidt’s and Epelbaum’s groups simultaneously advanced the force to next-to-next-to-next-to-leading order (N$^3$LO), introducing new contact terms, two-loop TPEs, three-pion exchanges and corrections, achieving precision comparable to the best phenomenological forces~\cite{Entem:2003ft,Epelbaum:2004fk}. Subsequent improvements included local regularization and uncertainty estimation~\cite{Epelbaum:2014efa}, with recent completion of local N$^3$LO in coordinate space~\cite{Saha:2022oep} and semi-local N$^4$LO$^+$ with semi-local regulator in momentum space~\cite{Reinert:2017usi}. The sixth-order (N$^5$LO) force remains incomplete, with only partial long-range contributions studied~\cite{Entem:2015xwa}.

It is worth noting that although the non-relativistic chiral nuclear force has achieved great success, it has also encountered some difficulties (both fundamental and empirical), such as relatively slow convergence~\cite{Epelbaum:2019zqc}, failure to satisfy renormalization-group invariance~\cite{Epelbaum:2018zli}, and inconsistencies in treating three-body forces~\cite{Epelbaum:2019zqc,Epelbaum:2019kcf,Epelbaum:2022cyo,Machleidt:2023jws}. To solve these problems, various attempts have been made, for example, developing chiral nuclear forces that do not explicitly contain pions~\cite{Bedaque:2002mn}, considering the $\Delta(1232)$~\cite{Ordonez:1993tn,Ordonez:1995rz,Kaiser:1998wa,Krebs:2007rh,Epelbaum:2008td,Piarulli:2014bda,Ekstrom:2017koy,Strohmeier:2020dkb,Nosyk:2021pxb,vanKolck:1994yi,Pandharipande:2005sx,Epelbaum:2007sq,Kaiser:2015yca,Krebs:2018jkc} and Roper~\cite{Xiao:2026pkb,Kaiser:2026lvl} contributions, utilizing renormalization group invariance to modify the Weinberg power counting~\cite{PhysRevC.72.054006,Birse:2005um,PhysRevLett.114.082502,PhysRevC.95.024001}, developing the relativistic chiral nuclear force~\cite{Ren:2016jna, Xiao:2018jot, Xiao:2020ozd, Wang:2021kos, Lu:2021gsb,Lu:2025ubc}, etc. This article primarily reviews recent efforts and progress in developing the relativistic chiral nuclear force.

The present review is organized as follows. First, we explain why a relativistic chiral nuclear force is necessary. We then show how to construct a nuclear force in the relativistic framework, reproduce scattering phase shifts and observables, and introduce recent efforts to construct higher-order relativistic chiral nuclear forces, study n-$d$ scattering, and apply relativistic chiral nuclear forces to studies of nuclear matter, finite nuclei, and hypernuclear systems. Finally, we summarize and present the prospects for future developments.

%%%%%%%%%%%%%%%%%%%%%%%%%%%%%%%%%%%%%%%%%%
\section{Why a relativistic chiral nuclear force}

Lorentz symmetry is one of the most essential symmetries in Nature. In principle, all physical laws should satisfy Lorentz symmetry. Of course, for non-relativistic systems,  it reduces to Galilean symmetry. Nonetheless, in the subatomic world, all systems should, in principle, satisfy Lorentz symmetry; that is, they should be studied using relativistic approaches. Relativity plays an important role in understanding many fine structures across different systems. For instance, in atomic systems, the different colors of Gold and silver are fully explained by relativistic effects~\cite{DESCLAUX20021}. In the nuclear system, the large spin-orbit interaction and the origin of pseudospin symmetry can be attributed to relativistic effects~\cite{Liang:2014dma}. In the single baryon system, the relativistic chiral effective field theory has been widely applied to the studies of baryon masses~\cite{Ren:2012aj,Ren:2013oaa,Ren:2013wxa,Ren:2013dzt, Ren:2016aeo,Chen:2024twu,Liang:2025cjd}, sigma terms~\cite{Ren:2014vea,Ren:2017fbv,Liang:2025adz}, magnetic moments ~\cite{Geng:2008mf,Liu:2018euh,Shi:2018rhk,Xiao:2018rvd,Shi:2021kmm}, meson-baryon scattering~\cite{Siemens:2016hdi,Lu:2022hwm}, hyperon weak radiative decay~\cite{Shi:2022dhw}, etc.) and shows a faster convergence than its non-relativistic counterpart. 

While the nonrelativistic chiral nuclear force has achieved considerable success, it still faces several unresolved issues that remain the focus of intense academic debate. The core challenge centers on power counting and the convergence of the chiral expansion, which are typically linked to renormalization-group invariance. The essence of the problem is that the Lippmann-Schwinger equation becomes non-renormalizable when truncated at a given order of the chiral expansion.~\cite{Epelbaum:2018zli}. See Ref.~\cite{Hammer:2019poc} for recent discussions. Attempts, such as treating the pion-exchange contribution perturbatively~\cite {Kaplan:2019znu}, have been proposed but suffer from unresolved convergence issues. In the covariant framework, it was found that the non-perturbative OPE is renormalizable~\cite{Baru:2019ndr}, but higher-order contributions still require perturbative treatment. Currently, there is no satisfactory solution to this problem. On the other hand, the convergence of the chiral expansion for the nonrelativistic chiral nuclear force is slow. It has been pointed out that accurately describing the neutron-deuteron scattering process requires at least the N$^4$LO chiral nuclear force~\cite{Epelbaum:2019zqc}.

The non-relativistic chiral nuclear force has become the standard microscopic input for \textit{ab initio}  nuclear physics. It has been widely applied in the studies of nuclear structure, nuclear reactions, nuclear matter, and nuclear astrophysics, achieving great success~\cite{Hebeler:2020ocj}. 
It is worth noting that to adapt to different many-body methods based on the chiral nuclear force, various optimized or softened nuclear forces have been developed, including NNLO$_{\text{sat}}$\cite{Ekstrom:2015rta}, NNLO$_{\text{opt}}$\cite{Ekstrom:2013kea}, and NNLO$_{\text{sim}}$\cite{Carlsson:2015vda}. 
%The differences lie in the methods used to determine the LECs of NN, 3N, and $\pi$N interactions, and in whether to introduce many-body observables as constraints. The technique of softening the nuclear force is based on the similarity renormalization group method~\cite{Jurgenson:2009qs,Roth:2011ar,Roth:2011vt}. The basic idea is to decouple the high-energy and low-energy parts of the Hamiltonian through a series of unitary transformations, thereby softening the nuclear force. Note that this will introduce equivalent three-body nuclear forces. 
However, Machleidt pointed out that many such methods are not truly \textit{ab initio} ~\cite{Machleidt:2023jws}. Among the two deemed truly \textit{ab initio}, namely, ``Hoppe''~\cite{Drischler:2017wtt} and ``Huether''~\cite{Huther:2019ont}, ``Hoppe'' can describe nuclear matter but not medium-mass nuclei, while ``Huether'' can describe medium-mass nuclei but not nuclear matter. 

Considering the problems encountered by the non-relativistic chiral nuclear force, especially the slow convergence and the failure to satisfy renormalization group invariance, we proposed to develop the relativistic chiral nuclear force based on the covariant baryon chiral effective field theory~\cite{Ren:2016jna,Li:2016mln,Lu:2021gsb}. This method follows the EOMS scheme to restore the power counting in the one-baryon system. Unlike the heavy baryon chiral effective field theory, it retains the complete Dirac spinor and Clifford algebra in the covariant chiral effective Lagrangian. Additionally, non-perturbative effects are accounted for by solving the relativistic scattering equation rather than the Lippmann-Schwinger equation. 

\section{How to construct a relativistic chiral nuclear force}

To construct relativistic chiral nuclear forces, we adopt the following key strategies:
\begin{enumerate}
    \item 
One first needs to construct the covariant chiral effective Lagrangians that satisfy chiral symmetry, parity, charge conjugation, Hermitian conjugation, and time-reversal invariance. Most importantly, the Lagrangians should be a Lorentz scalar. The chiral order of each building block in the covariant Lagrangians is listed in Table~\ref{tab1}

\begin{table}[H] 
%\small % Change table font size
\caption{Chiral orders of the nucleon bilinear structures, differential operators, Dirac matrices, and Levi-Civita tensors, as well as their properties under parity ($\mathcal{P}$), charge conjugation ($\mathcal{C}$), and Hermitian conjugation (h.c.) transformations.\label{tab1}}
%\isPreprints{\centering}{} % Only used for preprints
\begin{tabularx}{\textwidth}{CCCCCCCCC}
\toprule
Building Blocks&  $1$    &  $\gamma_5$    &    $\gamma_\mu$     &   $\gamma_5\gamma_\mu$   &   $\sigma_{\mu\nu}$  &   $\epsilon_{\mu\nu\rho\sigma}$ &   $\overleftrightarrow \partial_{\mu}$ & $\partial_{\mu}$\\
\midrule
$\mathcal{P}$   & $+$ & $-$ & $+$ & $-$ & $+$ & $-$ & $+$ & $+$\\
$\mathcal{C}$   & $+$ & $+$ & $-$ & $+$ & $-$ & $+$ & $-$ & $+$\\
  h.c.            & $+$ & $-$ & $+$ & $+$ & $+$ & $+$ & $-$ & $+$\\
\midrule
Chiral order   & $0$ & $1$ & $0$ & $0$ & $0$ & $-$ & $0$ & $1$\\
\bottomrule
\end{tabularx}

\end{table}

Note that one should adopt the complete form of Dirac spinors and Clifford algebra instead of non-relativistic wave functions and Pauli matrices. The Dirac spinor is given by:
$$u(p, s)=N_{p}\left(\frac{\sigma \cdot p}{\epsilon_{p}}\right) \chi_{s}, \quad N_{p}=\sqrt{\frac{\epsilon_{p}}{2 M_{N}}}.$$

\item Employing a covariant power counting instead of the traditional Weinberg power counting, which ensures the Lorentz invariance of the interaction vertices and the scattering amplitude. When chiral effective field theory is applied to baryon systems, the power counting rule developed based on naive dimensional analysis is broken due to the non-zero baryon mass in the chiral limit. This is because the non-zero baryon mass makes the covariant derivative of the baryon field no longer a strictly small quantity~\cite{Scherer:2002tk}. To address this problem, one extends the covariant EOMS scheme~\cite{Fuchs:2003qc,Geng:2013xn}, which has been widely adopted in the one-baryon system, to nucleon-nucleon interactions~\cite{Lu:2025syk}.

\item Solving the covariant scattering equation instead of the non-relativistic Lippmann-Schwinger equation. The covariant scattering equation is expressed as:
        $$\mathcal{T}(p', p|W) = \mathcal{A}(p', p|W) + \int \frac{d^4k}{(2\pi)^4} \mathcal{A}(p', k|W) G(k|W) \mathcal{T}(k, p|W)$$
        where $$G(k | W)=\frac{i}{\left[\gamma^{\mu}(W+k)_{\mu}-m_{N}+i \epsilon\right]\left[\gamma^{\mu}(W-k)_{\mu}-m_{N}+i \epsilon\right]}$$ is the relativistic propagator of nucleons.

However, when the complex, realistic nuclear force is taken as the kernel, the aforementioned 4-dimensional Bethe-Salpeter equation is difficult to solve strictly. Various strategies have been proposed to conduct a 3-dimensional reduction, which result in, e.g., the Blankenbecler-Sugar equation~\cite{Blankenbecler:1965gx} or the Kadyshevsky equation~\cite{Kadyshevsky:1967rs} for practical applications.

\end{enumerate}

\section{The first high-precision relativistic chiral nuclear force}

We have constructed the first high-precision relativistic chiral NN force up to NNLO~\cite{Lu:2021gsb}. Up to this order, the complete relativistic chiral nuclear force consists of the following contributions
\begin{equation}\label{NNForce}  V=V_{\mathrm{CT}}^{\mathrm{LO}}+V_{\mathrm{CT}}^{\mathrm{NLO}}+V_{\mathrm{OPE}}+V_{\mathrm{TPE}}^{\mathrm{NLO}}+V_{\mathrm{TPE}}^{\mathrm{NNLO}}-V_{\mathrm{IOPE}},
\end{equation}
which includes the contact interactions at LO [$\mathcal{O}(p^0)$] and NLO [$\mathcal{O}(p^2)$], the OPE terms, the leading order and next-to-leading order TPE contributions. The last term subtracts the double-counted OPE contributions.

To be more explicit, the kernel involves three key inputs:
\begin{enumerate}
\item Contact Lagrangians\cite{Xiao:2018jot}, which describe the short-range interactions between nucleons.

\item Meson-baryon vertices~\cite{Lu:2018zof,Lu:2022hwm}, which govern the interactions between nucleons through meson exchanges.

\item Two-meson exchanges~\cite{Xiao:2020ozd,Wang:2021kos}, which contribute to the medium-range part of the nuclear force.
\end{enumerate}

\begin{figure}[H]
%\isPreprints{}{% This command is only used for ``preprints''.
\begin{adjustwidth}{-\extralength}{0cm}
\centering
\subfloat[\centering]{\includegraphics[width=5.0cm]{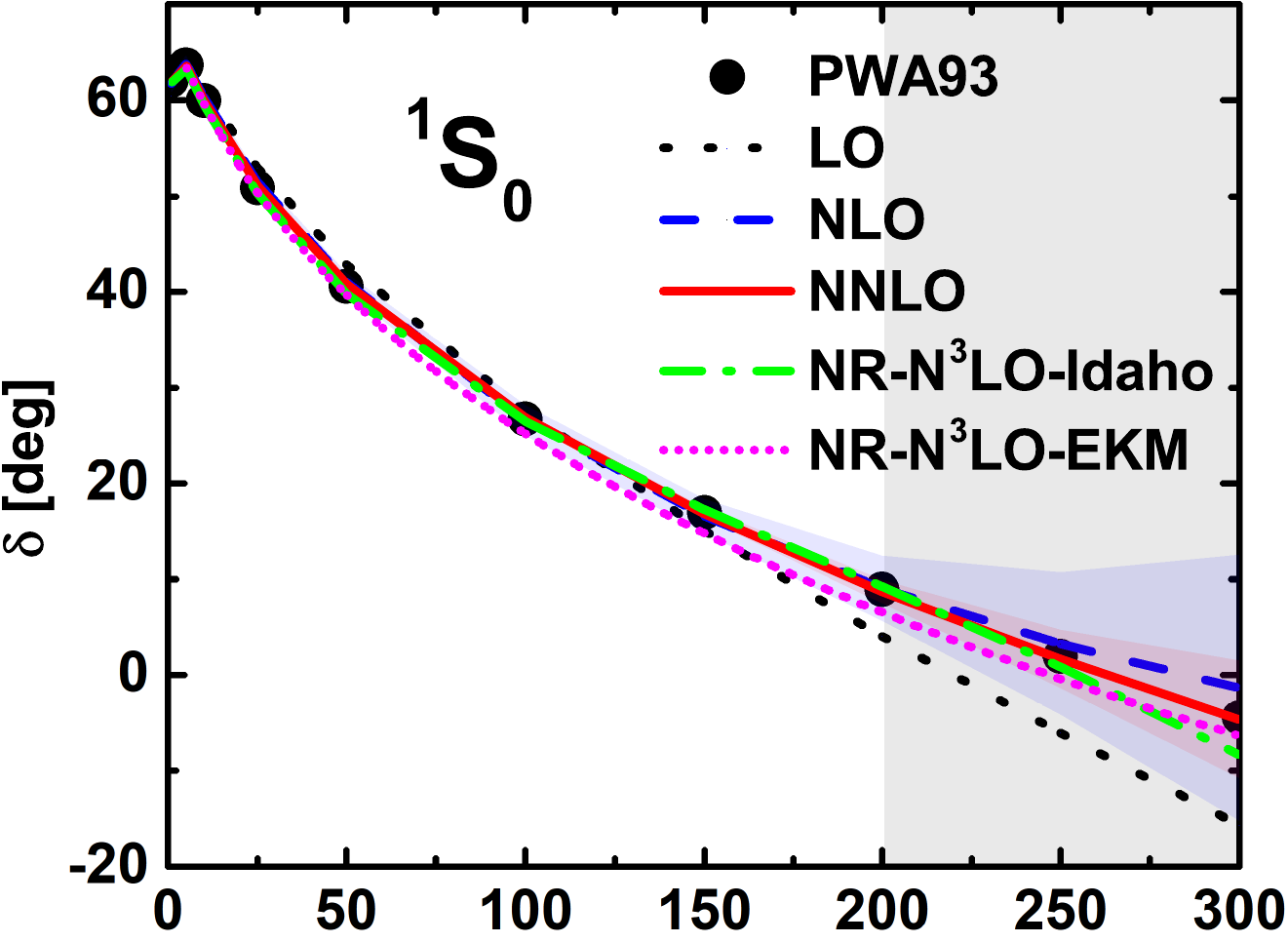}}
%\hfill
\subfloat[\centering]{\includegraphics[width=5.0cm]{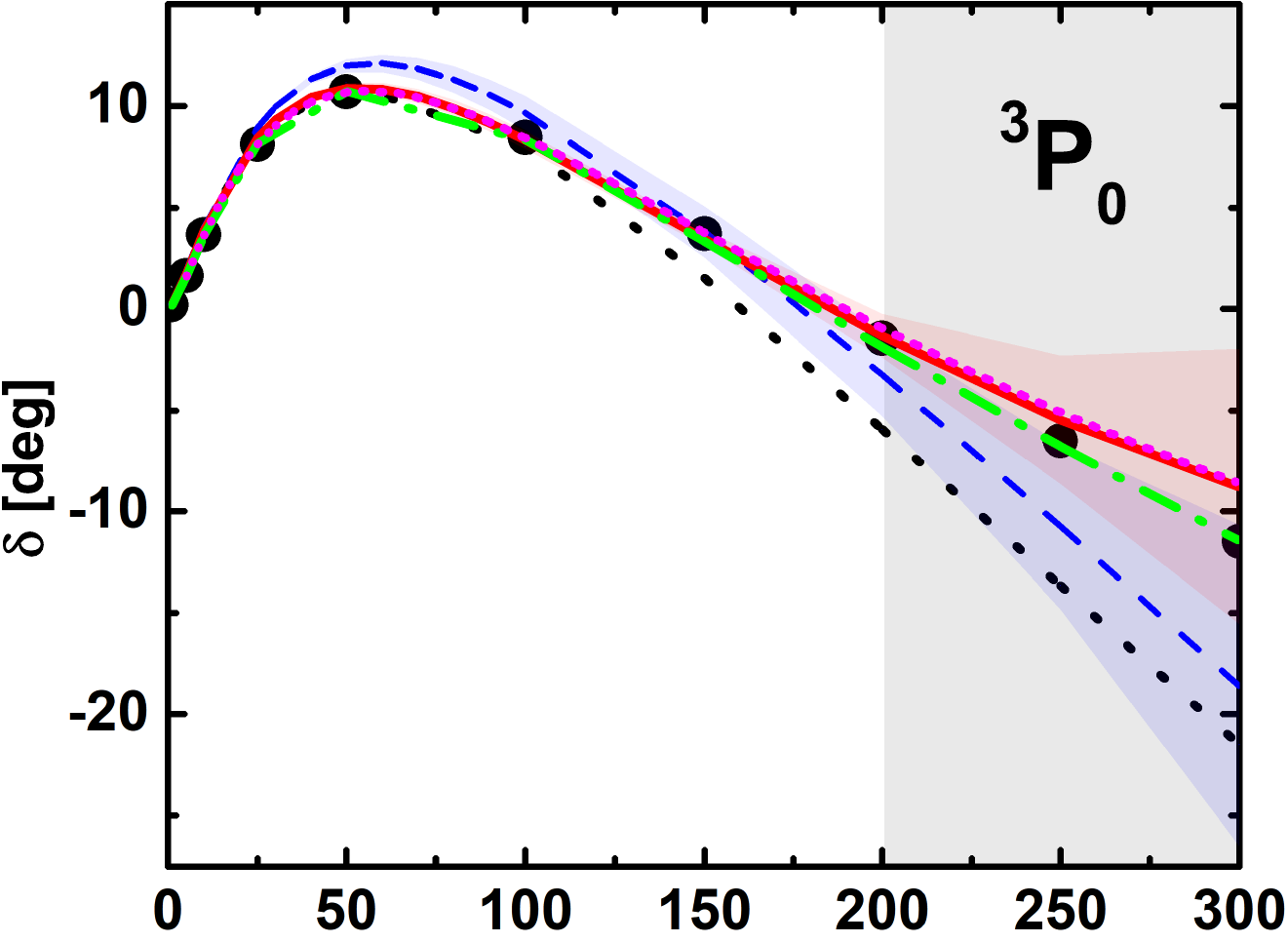}}
\subfloat[\centering]{\includegraphics[width=5.0cm]{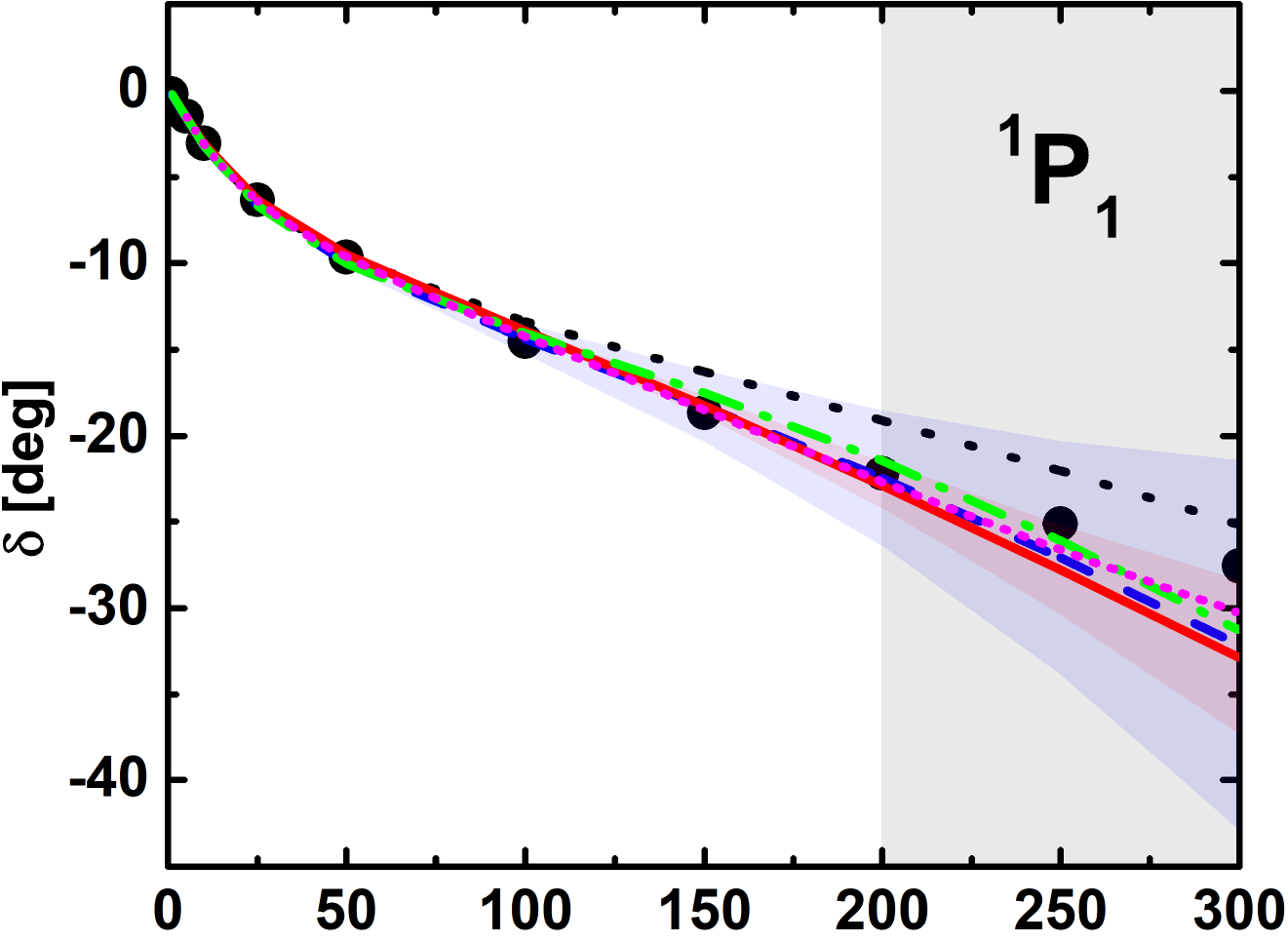}}\\
\subfloat[\centering]{\includegraphics[width=5.0cm]{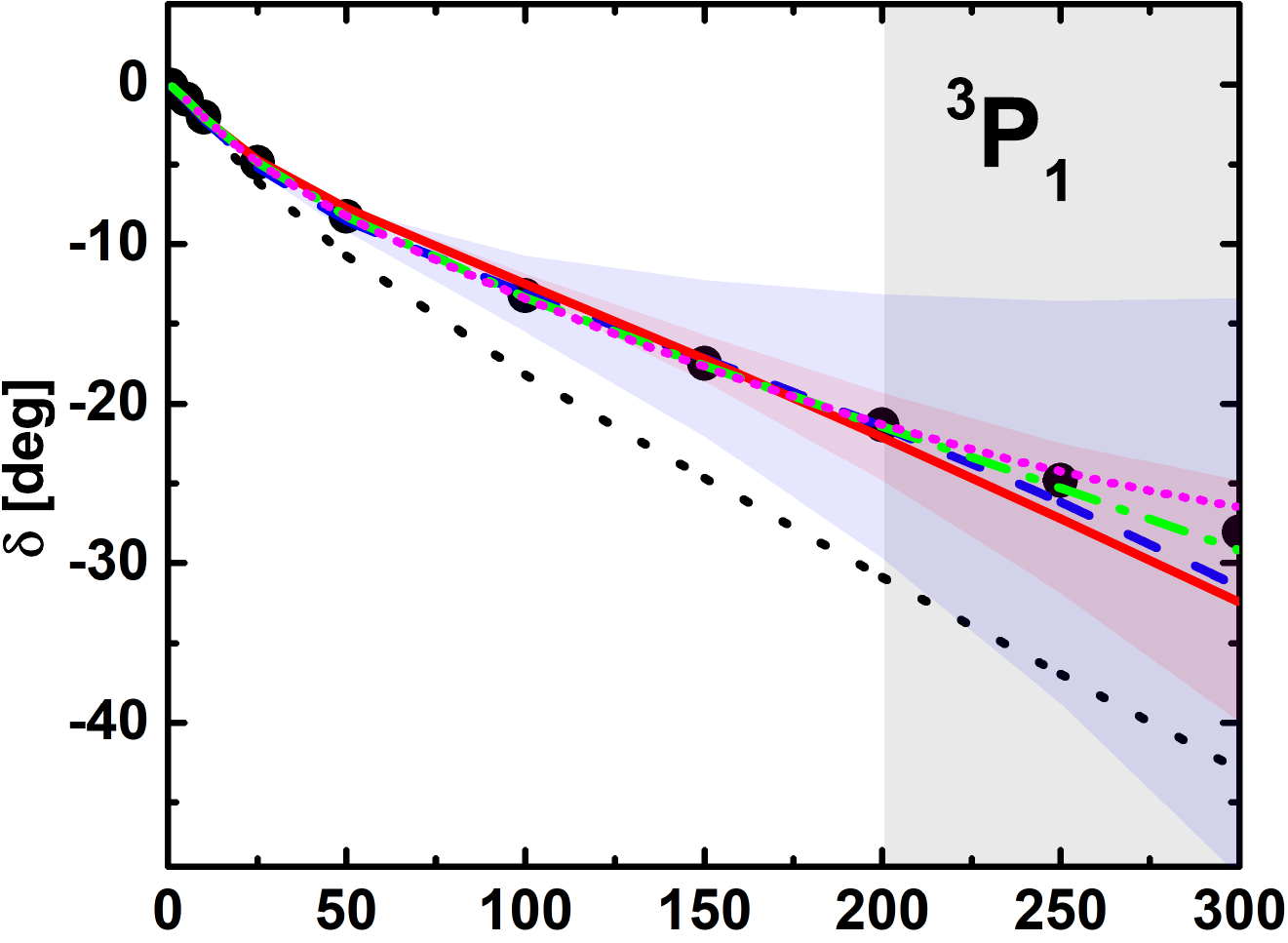}}
%\hfill
\subfloat[\centering]{\includegraphics[width=5.0cm]{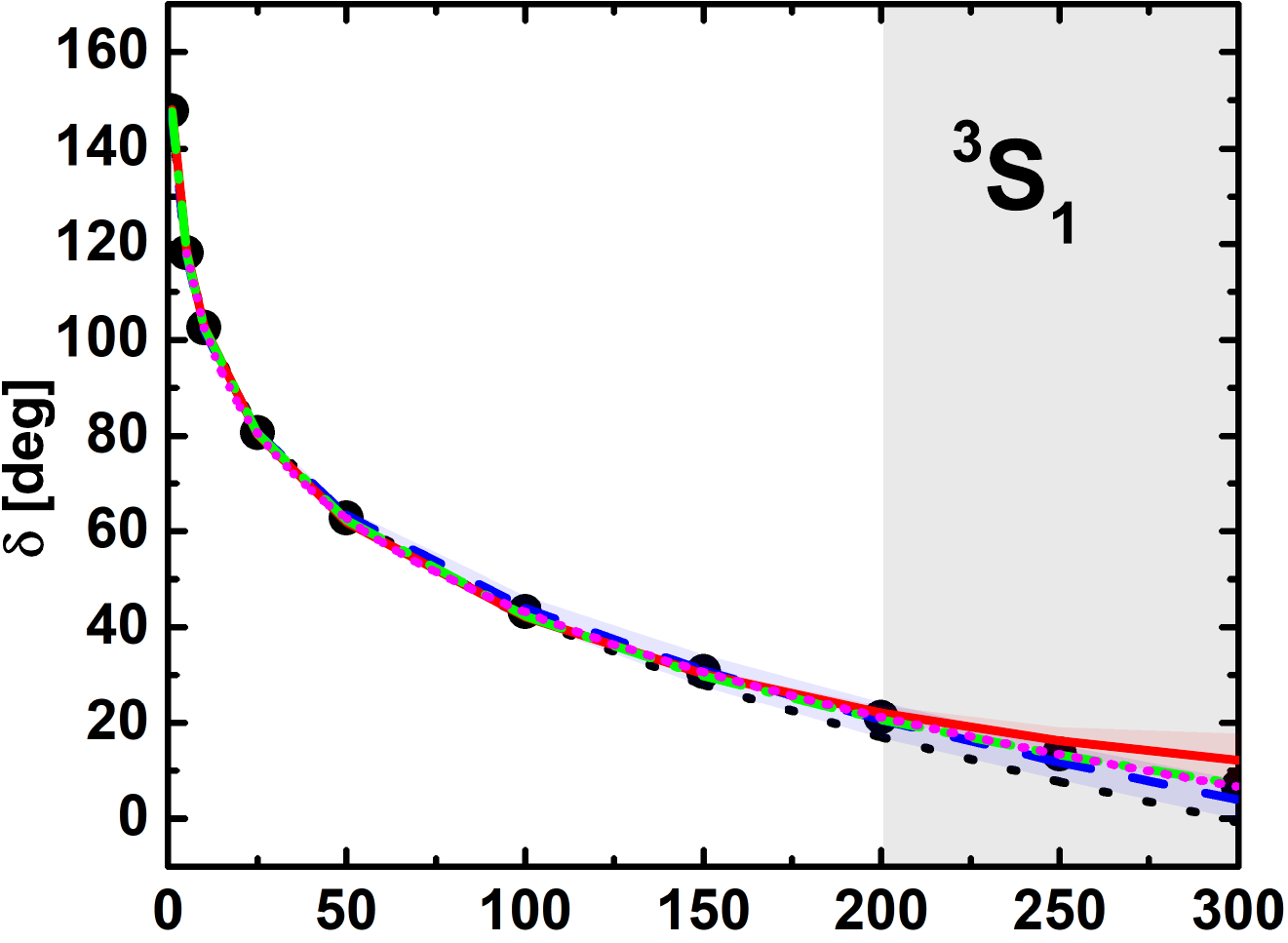}}
\subfloat[\centering]{\includegraphics[width=5.0cm]{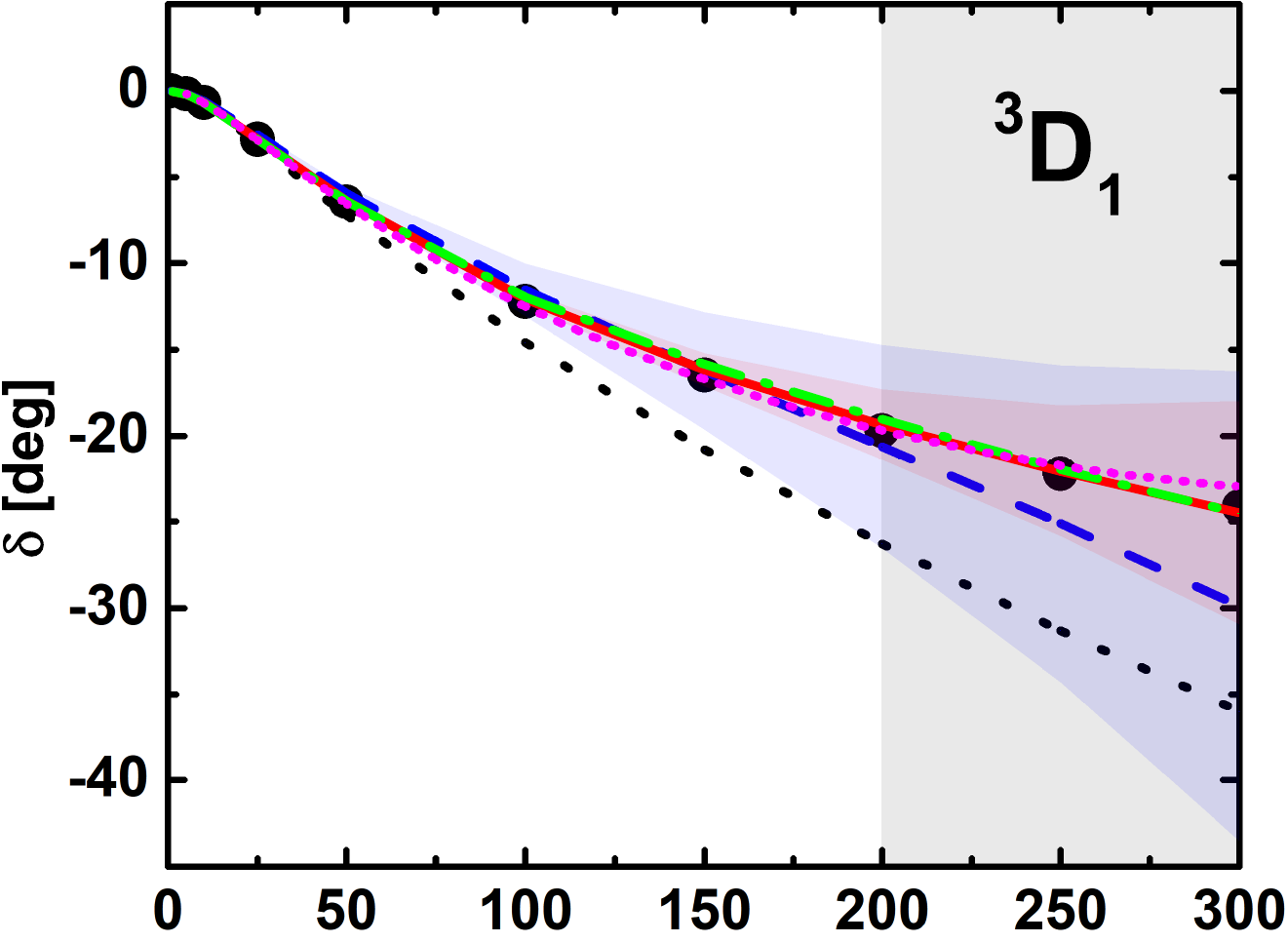}}\\
\subfloat[\centering]{\includegraphics[width=5.0cm]{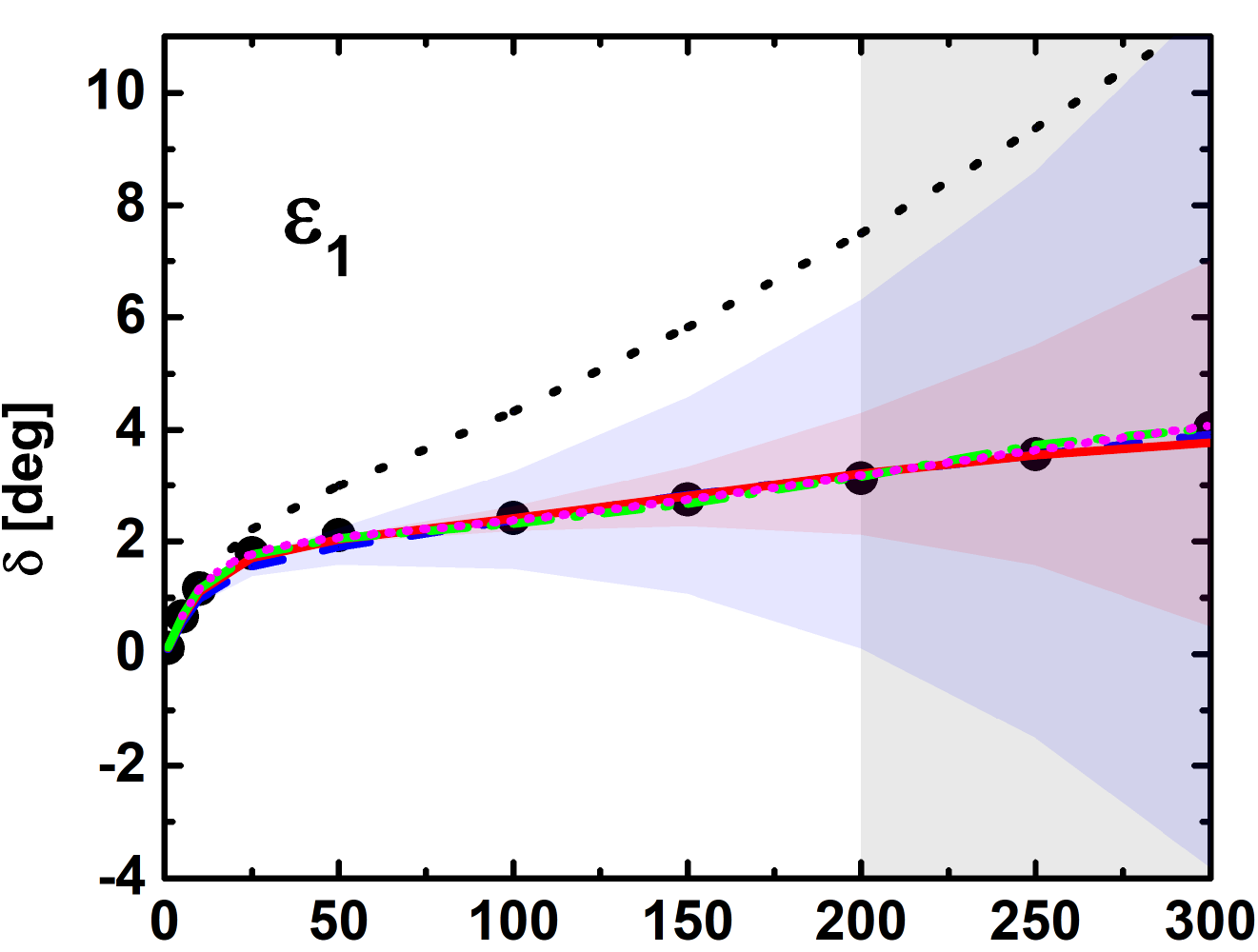}}
%\hfill
\subfloat[\centering]{\includegraphics[width=5.0cm]{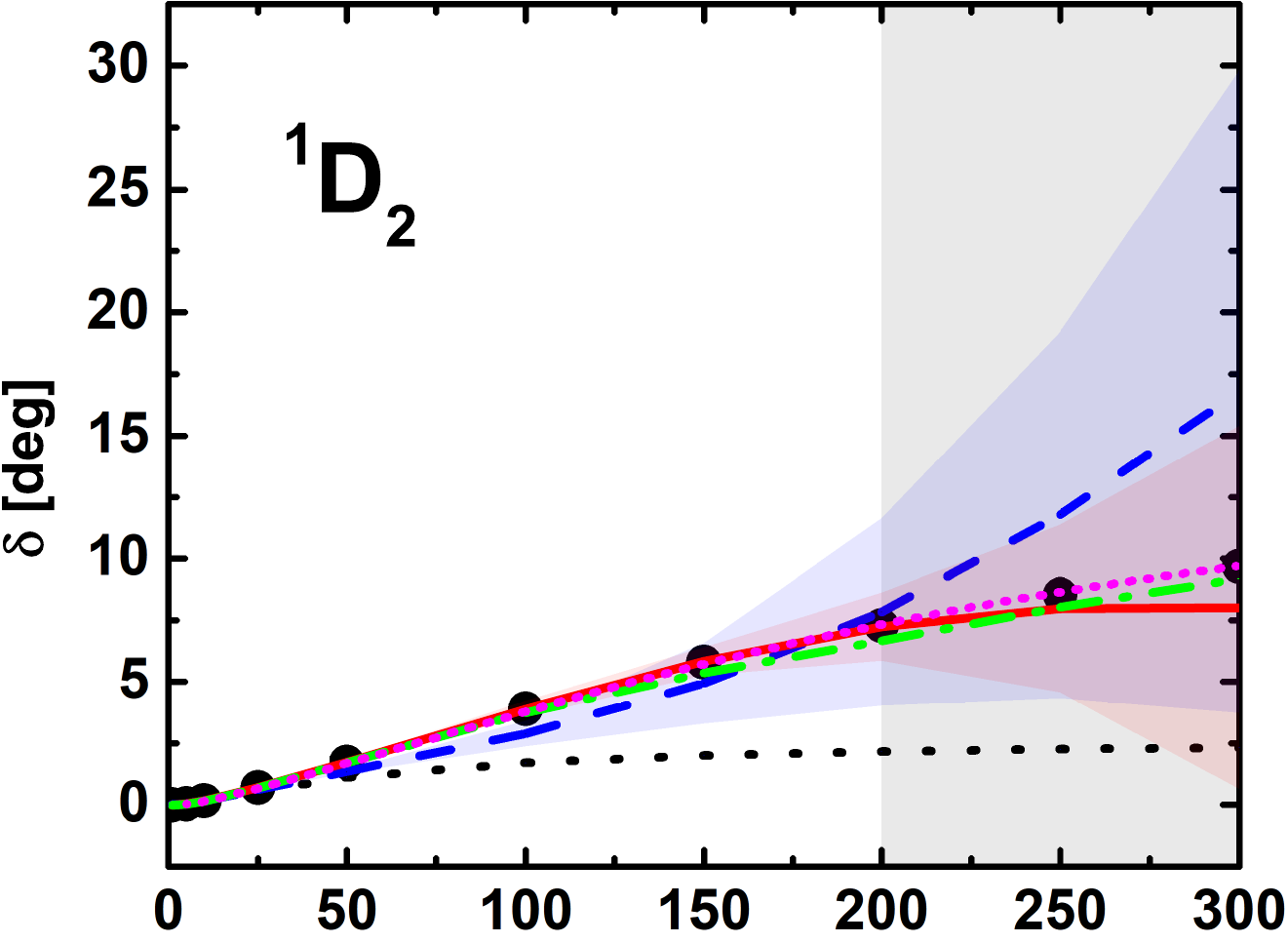}}
\subfloat[\centering]{\includegraphics[width=5.0cm]{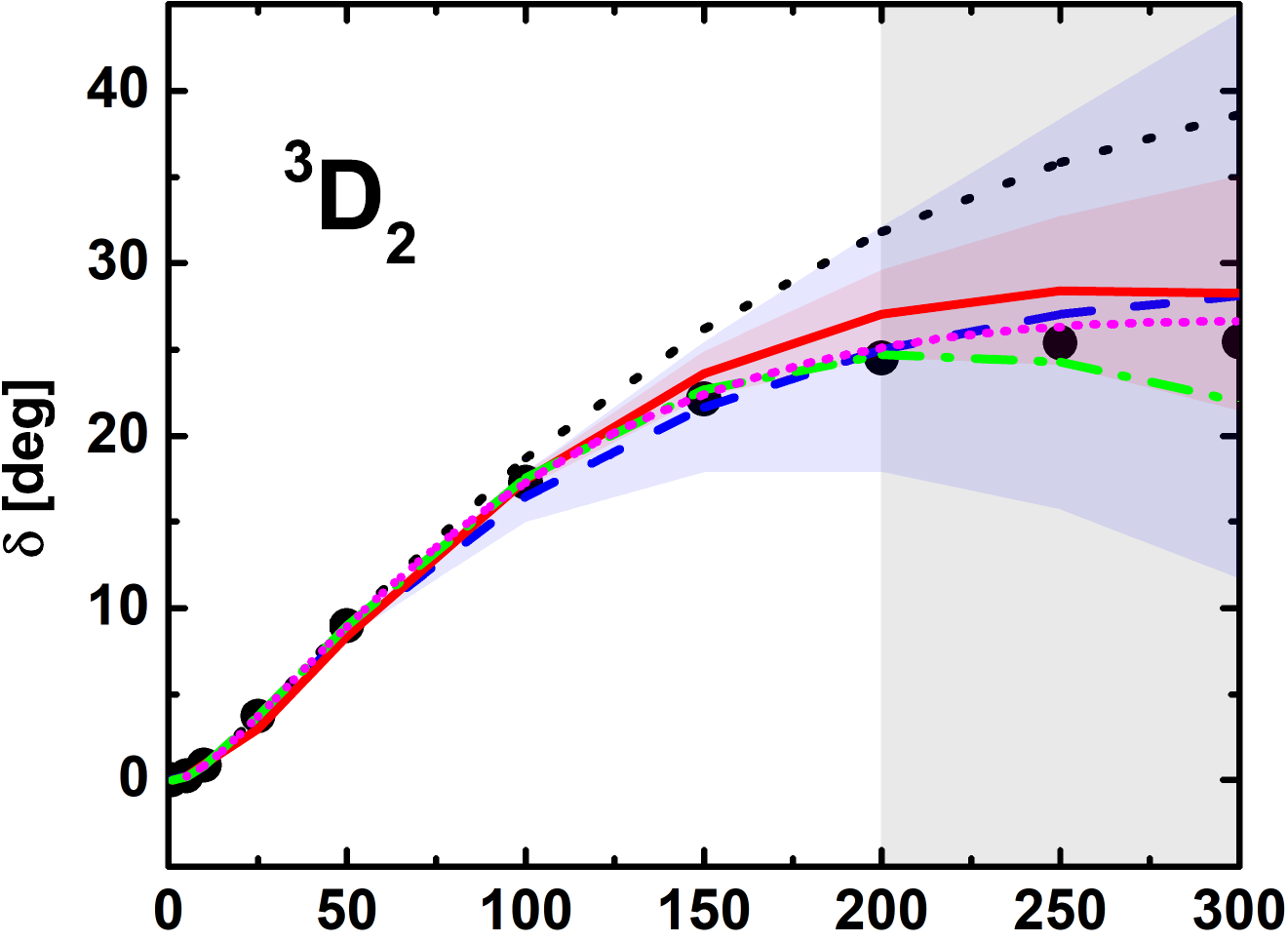}}\\
\subfloat[\centering]{\includegraphics[width=5.0cm]{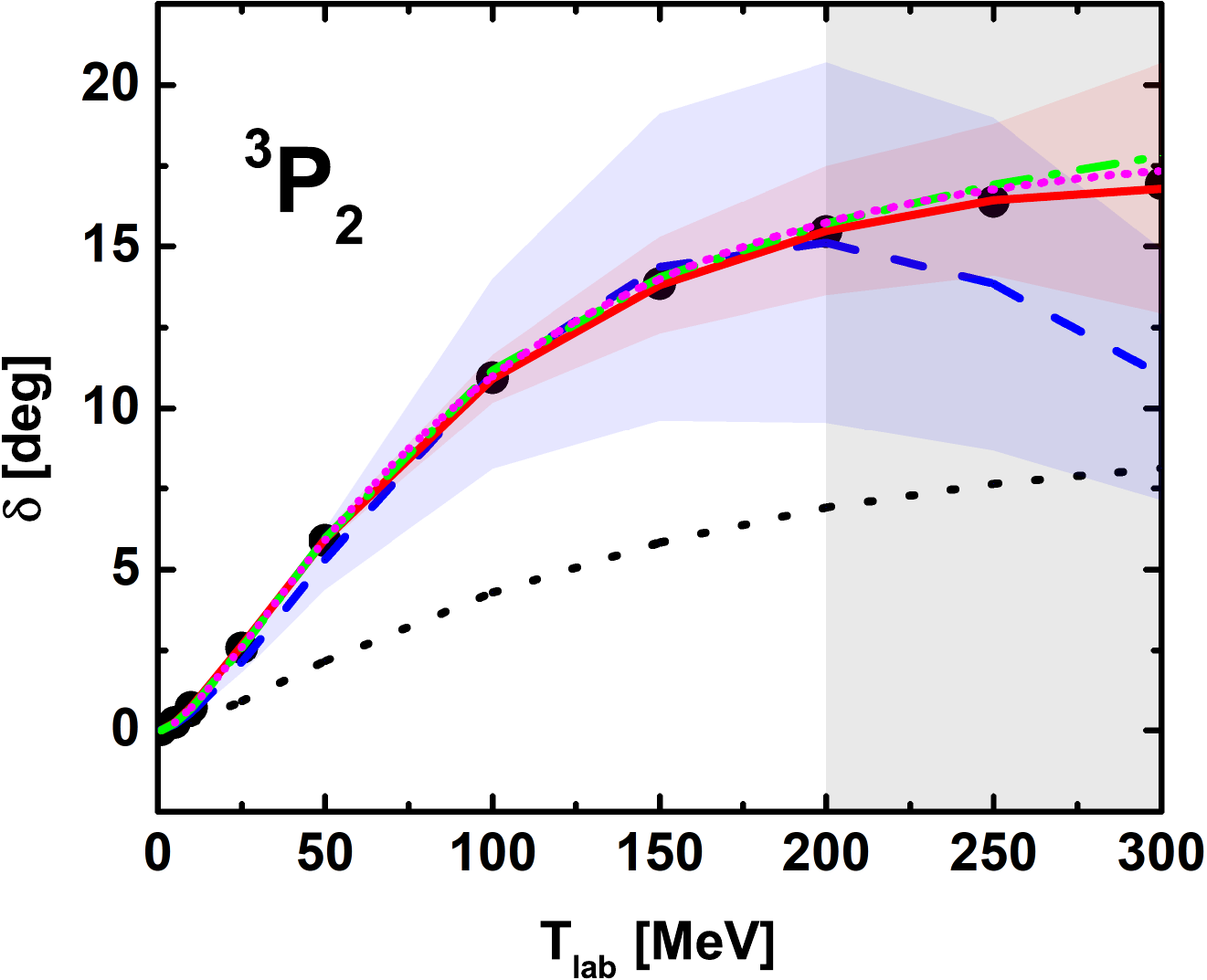}}
%\hfill
\subfloat[\centering]{\includegraphics[width=5.0cm]{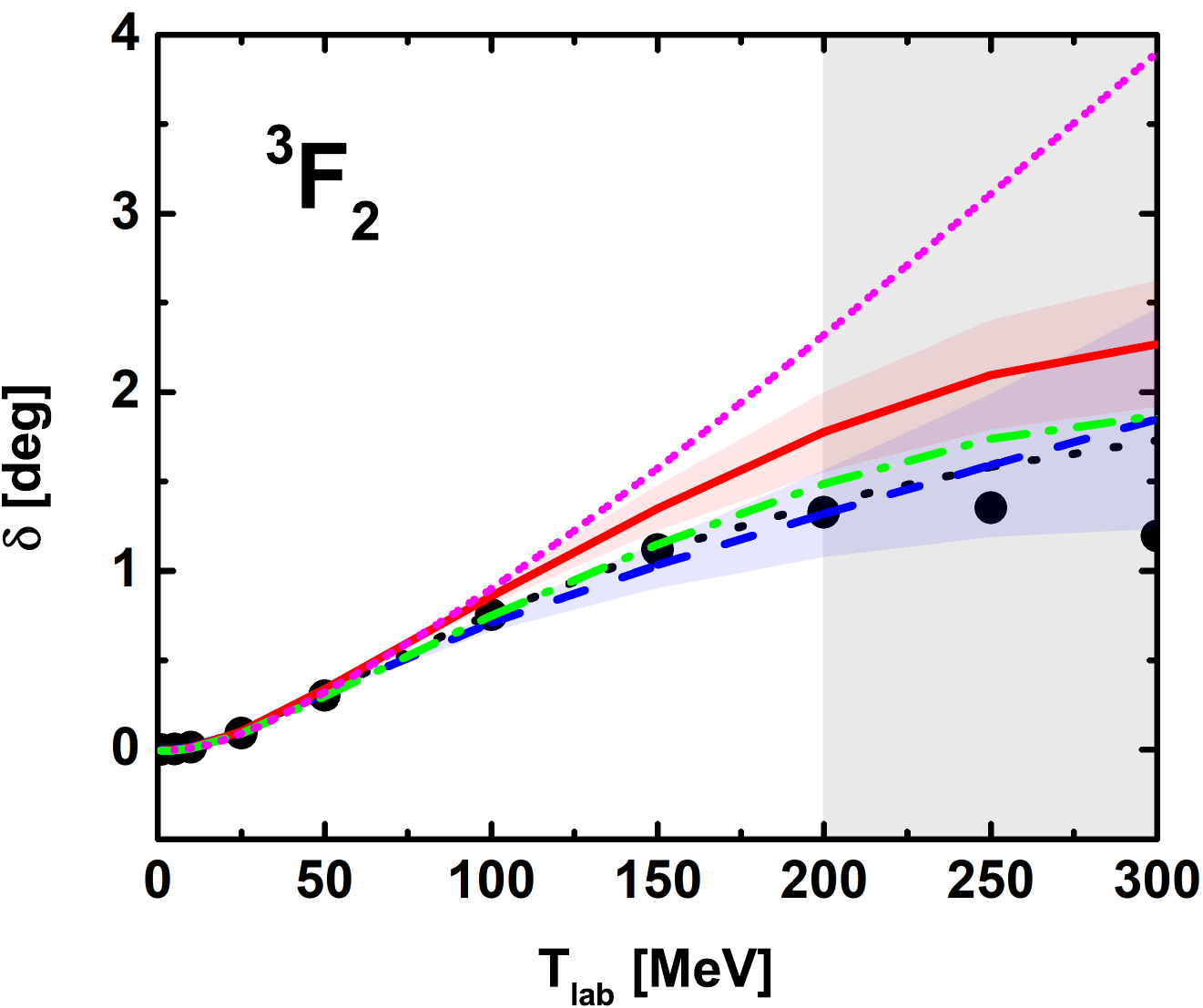}}
\subfloat[\centering]{\includegraphics[width=5.0cm]{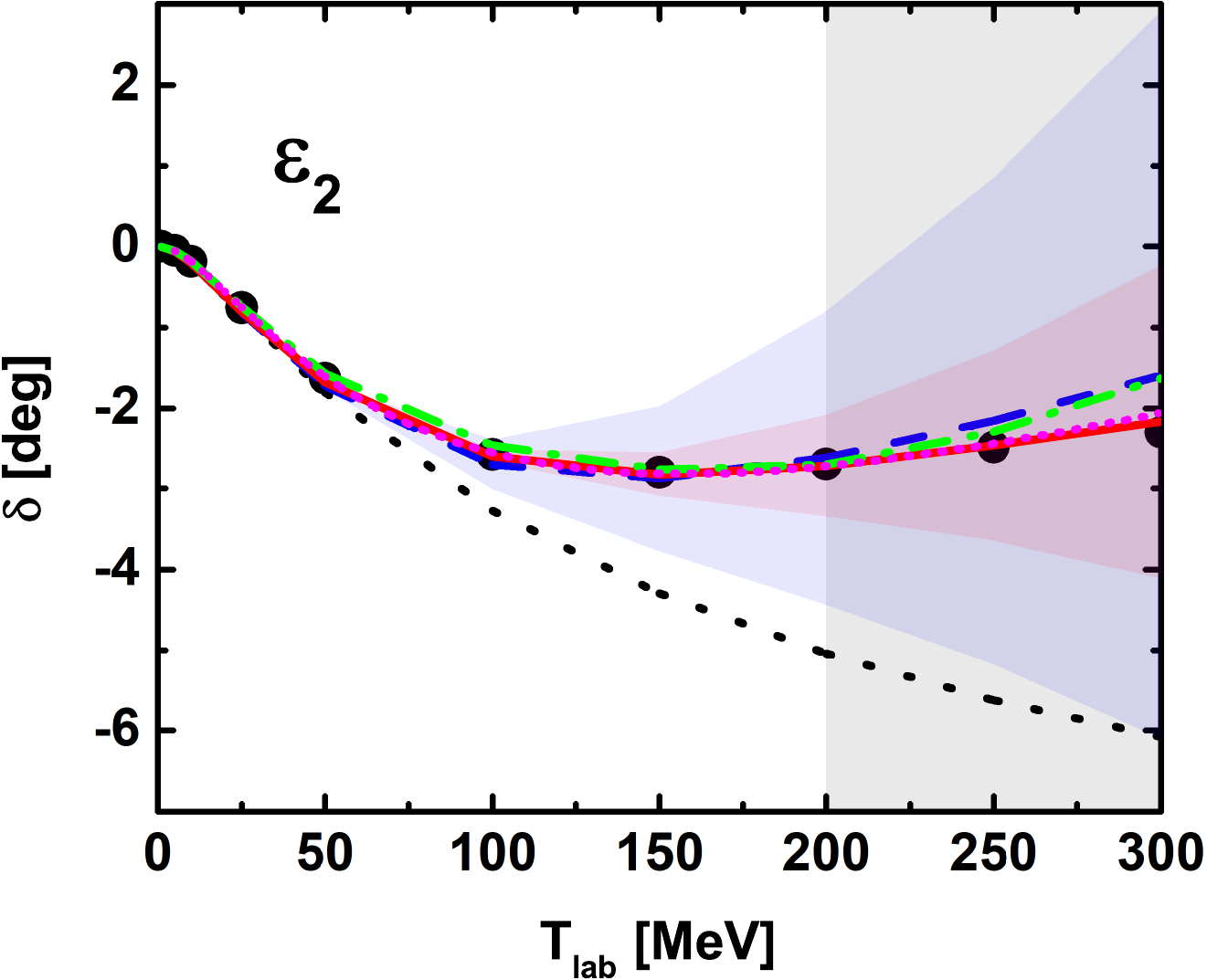}}\\
%\isPreprints{}{% This command is only used for ``preprints''.
\end{adjustwidth}
\caption{Scattering phase shifts of the $J\leq 2$ partial waves provided by different nuclear forces. The red solid lines result from our NNLO relativistic chiral nuclear force, and the momentum cutoff is set at $\Lambda=0.9$ GeV. The blue dashed lines result from the NLO relativistic chiral nuclear force, and the momentum cutoff is set at $\Lambda=0.6$ GeV. The corresponding shaded intervals are the theoretical uncertainties at the 68$\%$ confidence level. For comparison, we also present the results of the relativistic leading order (black dotted line, momentum cutoff $\Lambda=0.6$ GeV) and two non-relativistic N$^3$LO chiral nuclear forces, namely NR-N$^3$LO-Idaho ($\Lambda=0.5$ GeV, green dot-dashed line)~\cite{Entem:2003ft,Machleidt:2011zz}, and NR-N$^3$LO-EKM ($\Lambda=0.9$ fm, purple short dot-dashed line)~\cite{Epelbaum:2014efa,Epelbaum:2014sza}. The black solid points are the Nijmegen partial-wave analysis results~\cite{Stoks:1993tb}. Taken from Ref.~\cite{Lu:2021gsb}.\label{fig1}}
\end{figure}

\begin{figure}[H]
%\isPreprints{}{% This command is only used for ``preprints''.
\begin{adjustwidth}{-\extralength}{0cm}
\centering
\subfloat[\centering]{\includegraphics[width=5.0cm]{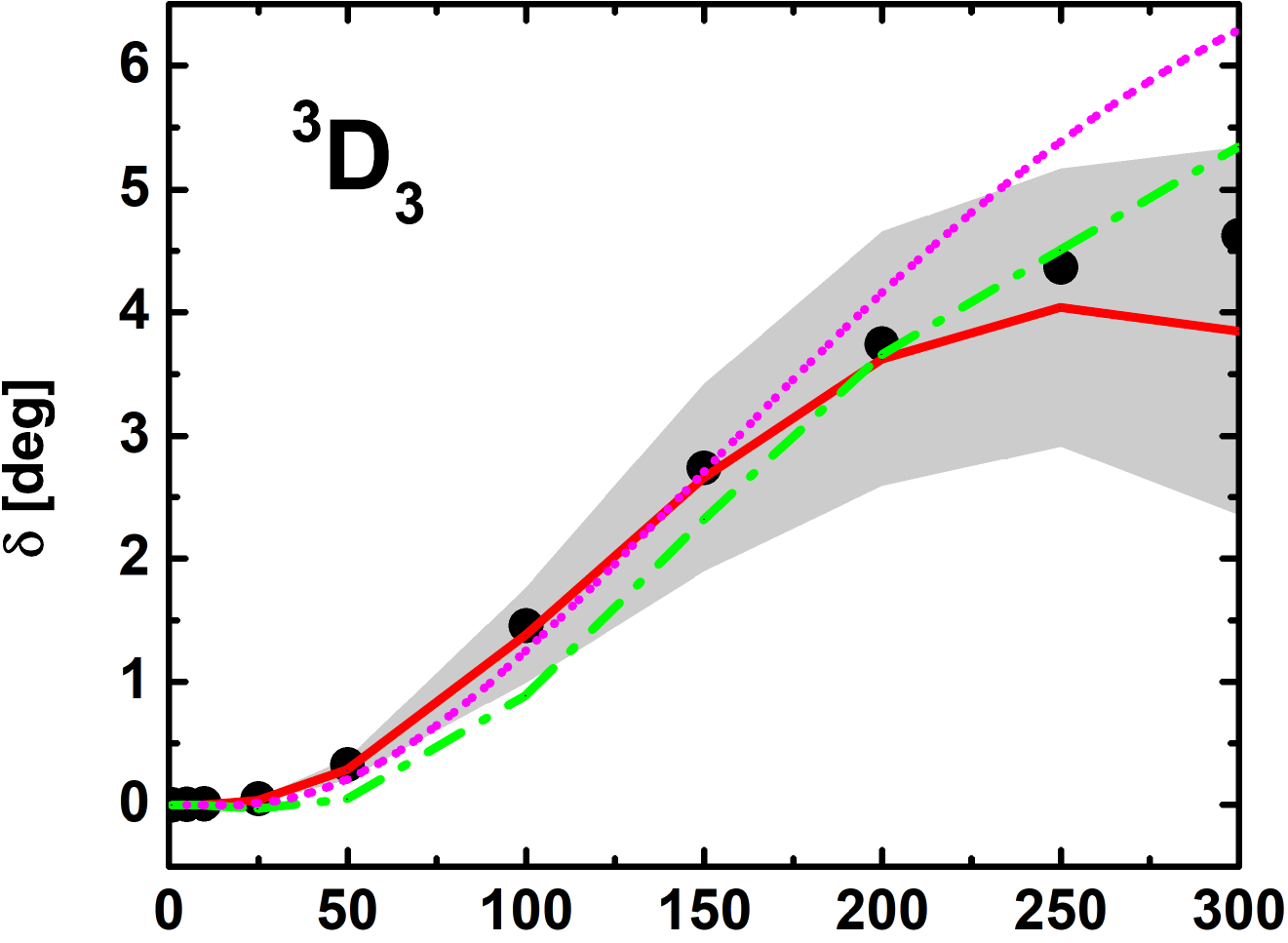}}
%\hfill
\subfloat[\centering]{\includegraphics[width=5.0cm]{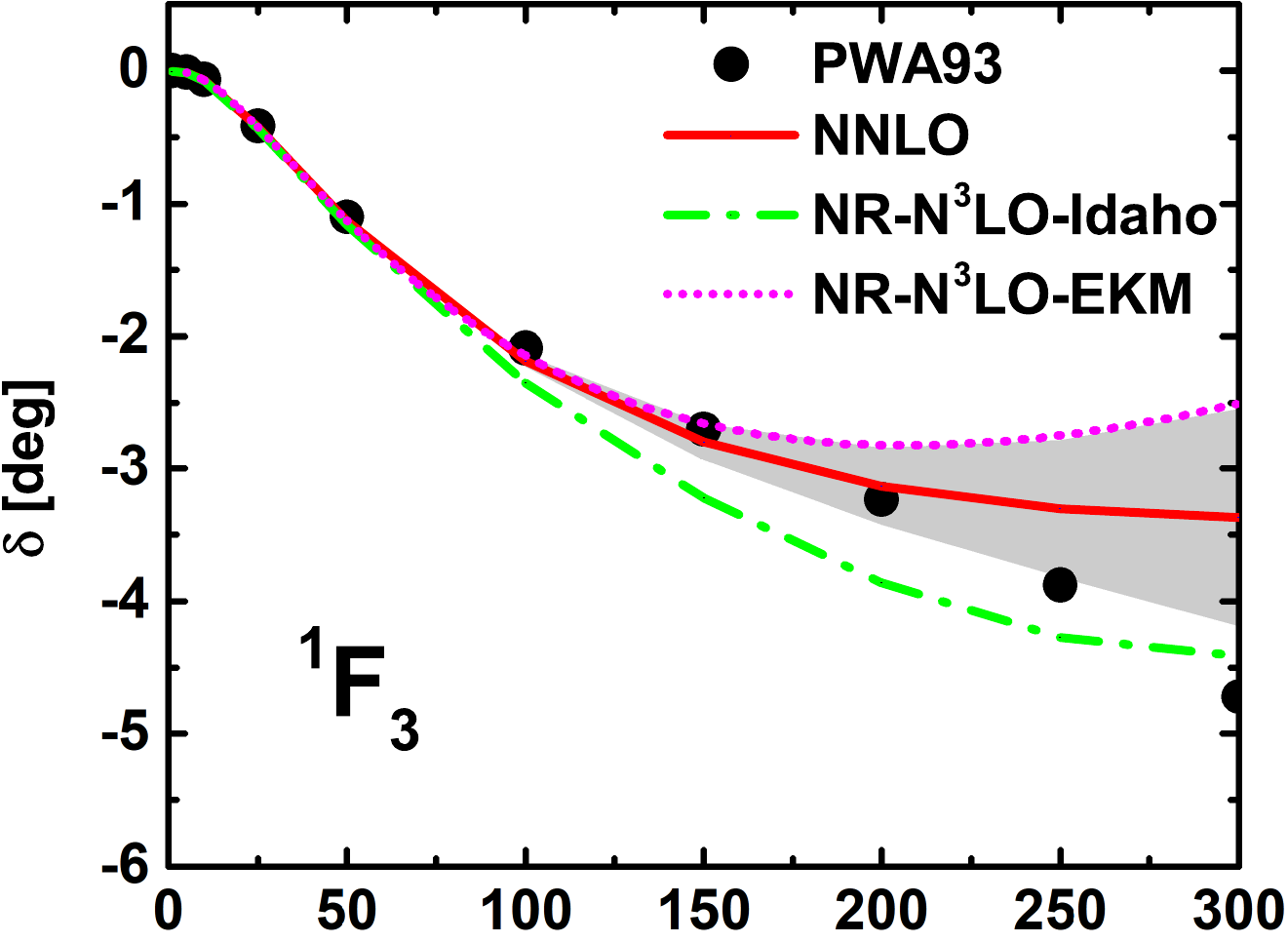}}
\subfloat[\centering]{\includegraphics[width=5.0cm]{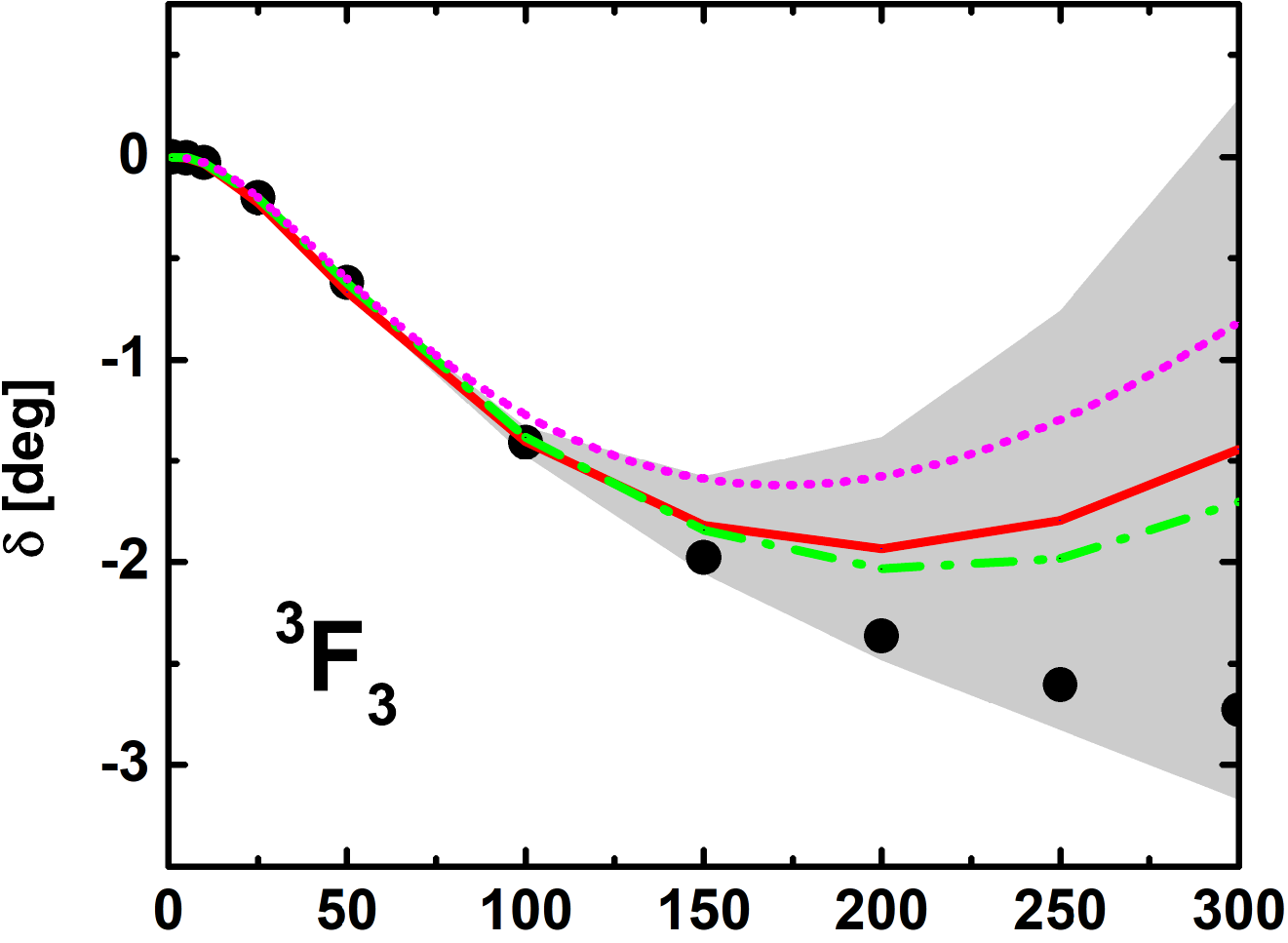}}\\
\subfloat[\centering]{\includegraphics[width=5.0cm]{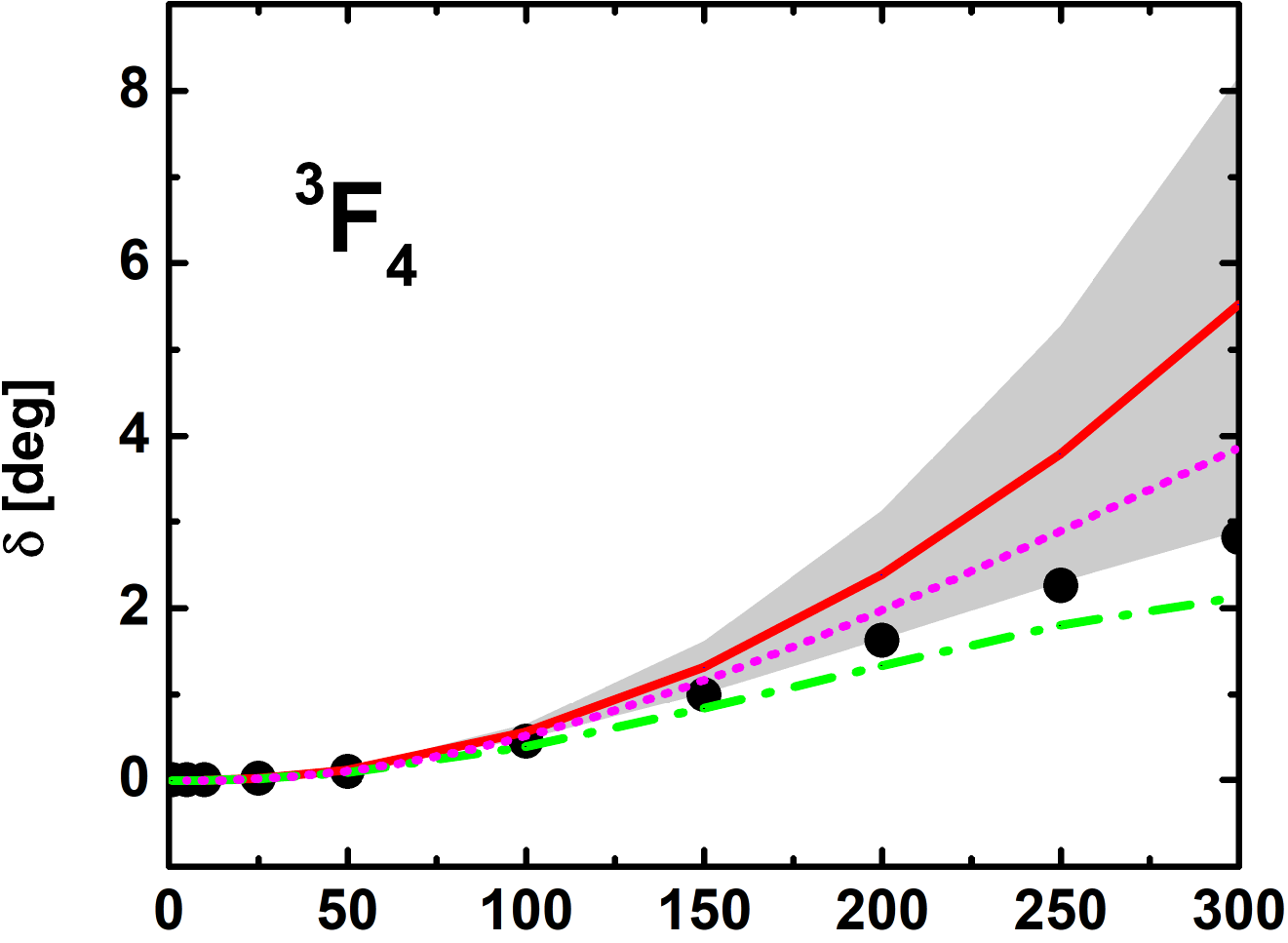}}
%\hfill
\subfloat[\centering]{\includegraphics[width=5.0cm]{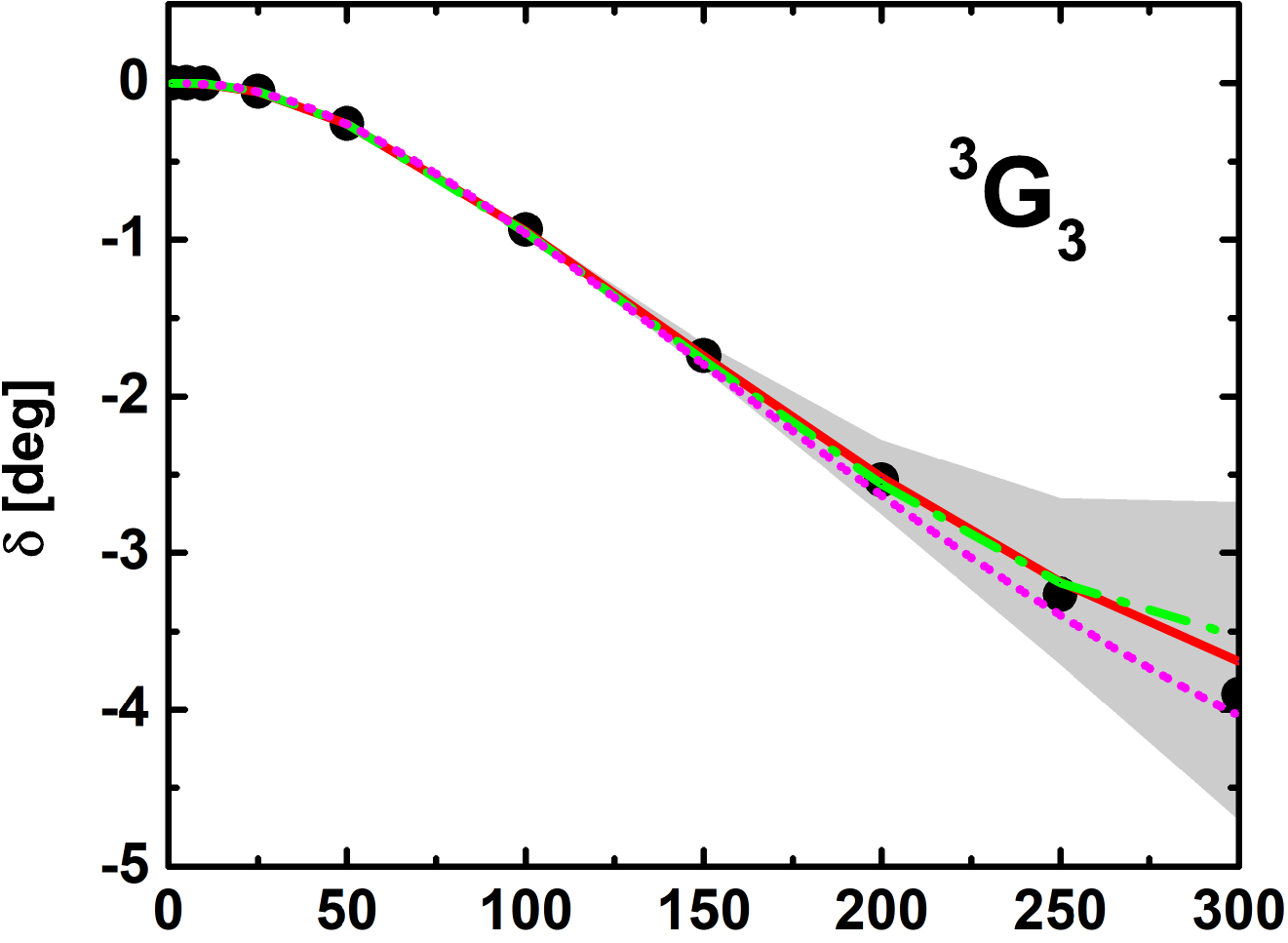}}
\subfloat[\centering]{\includegraphics[width=5.0cm]{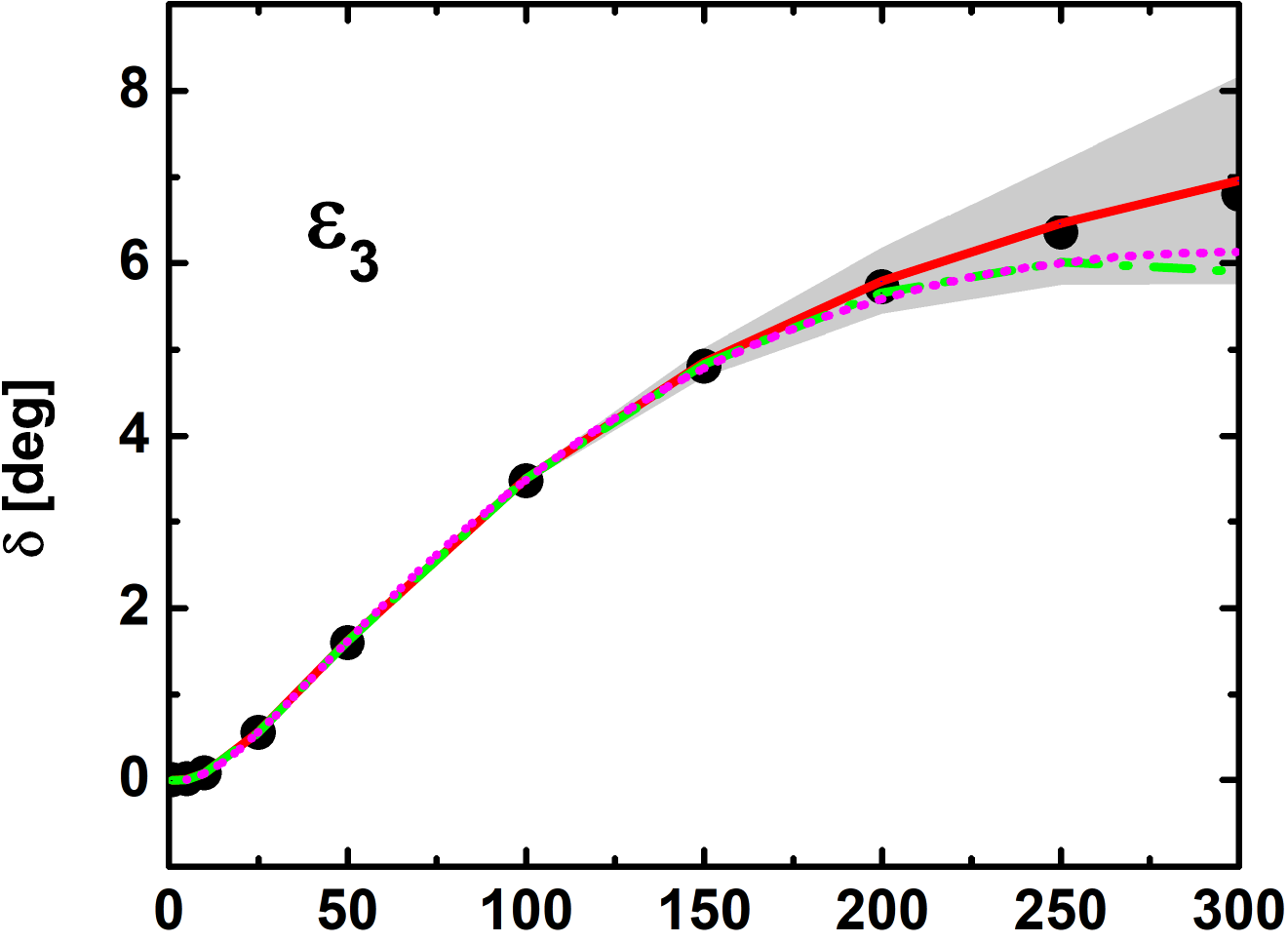}}\\
\subfloat[\centering]{\includegraphics[width=5.0cm]{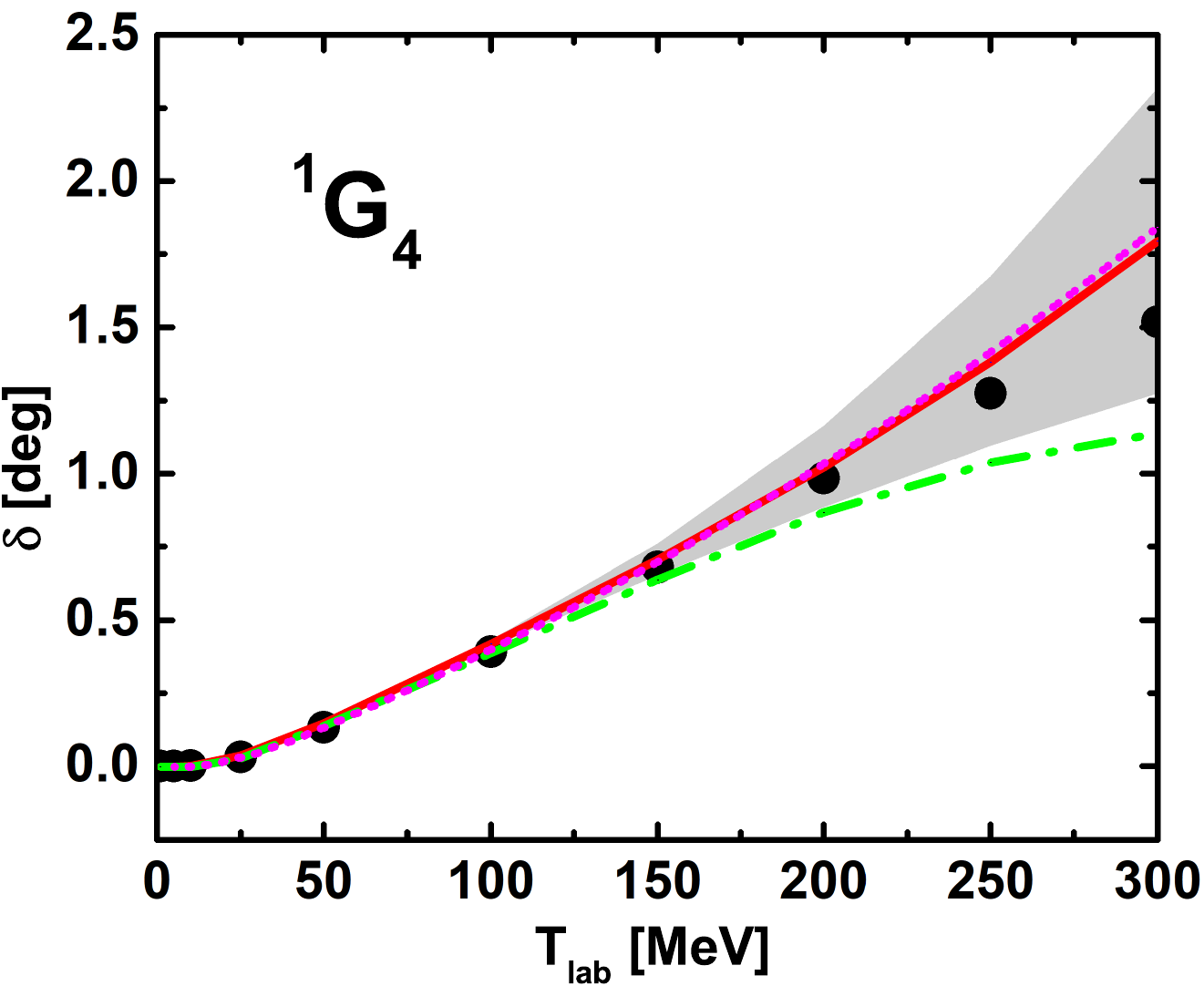}}
%\hfill
\subfloat[\centering]{\includegraphics[width=5.0cm]{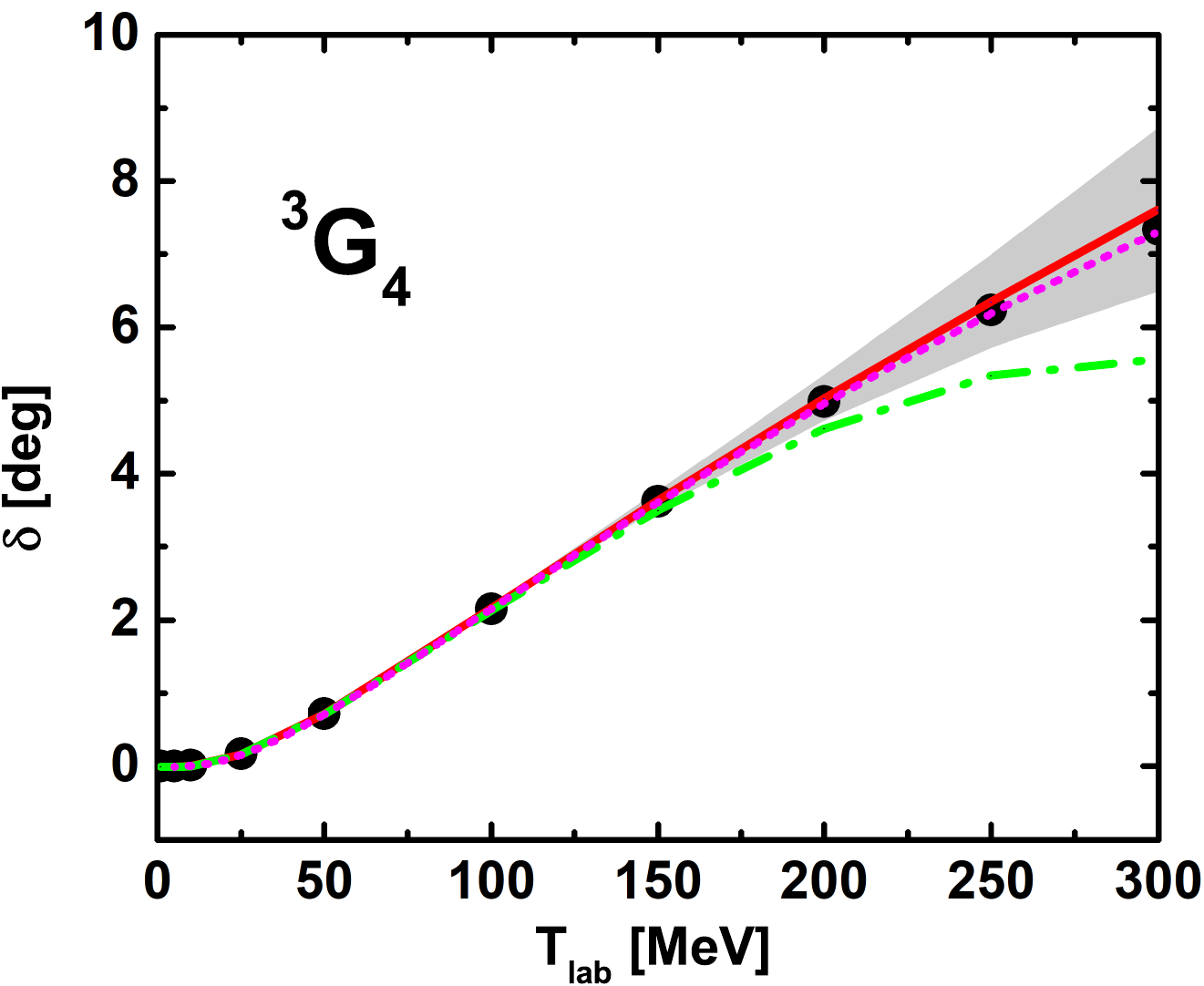}}\\
%\isPreprints{}{% This command is only used for ``preprints''.
\end{adjustwidth}
\caption{Same as Figure~\ref{fig1} but for peripheral partial waves with $J \leq 4$ and $L \leq 4$. Note that for these partial waves, the chiral results are pure
predictions without any free LECs. Taken from Ref.~\cite{Lu:2021gsb}.\label{fig2}}
\end{figure}

At this order, one has 19 LECs. They are fitted to the phase-shift data from PWA93. The fitting details are provided in Ref.~\cite{Lu:2021gsb}. 
The fit results for partial waves with \(J \leq 2\) are shown in Figure~\ref{fig1}. The theoretical uncertainty at the 68$\%$ confidence level is estimated by the Bayesian method~\cite{Furnstahl:2015rha,Melendez:2017phj,Melendez:2019izc}. For comparison, we also present the results obtained by the Weinberg N$^3$LO with different regularization schemes, denoted by ``Idaho" \cite{Entem:2003ft,Machleidt:2011zz} and ``EKM" \cite{Epelbaum:2014efa,Epelbaum:2014sza}, respectively. Within the relativistic framework, the NLO and NNLO descriptions of the phase shifts are in very good agreement with the experimental data up to $T_\mathrm{lab}=200$ MeV. They are also similar to the results of the Weinberg N$^3$LO chiral forces. In particular, the relativistic NLO phase shifts almost overlap with the NNLO ones, which better agrees with the experimental data. For higher partial waves shown in Figure \ref{fig2}, the relativistic treatment shows clear advantages, with a $\chi^2$ value of 0.99 for NNLO—lower than the 3.00 of NR-N$^3$LO-EKM—demonstrating faster convergence. 

Compared with non-relativistic chiral nuclear forces, the relativistic ones exhibit three prominent features: better renormalizability~\cite{Ren:2016jna}, faster convergence~\cite{Lu:2021gsb,Lu:2025ubc}, and more natural descriptions of nuclear interactions~\cite{Ren:2017yvw}—all of which may play a role in solving the challenges faced by non-relativistic theories. As shown later, relativistic chiral forces at LO can already capture key nuclear dynamics with only four LECs, without recourse to 3N forces~\cite{Zou:2023quo,Zou:2025dao,Shen:2025iue}.

% Example of a figure that spans the whole page width and has subfigures. The same concept works for tables, too.

% Example of a figure that spans the whole page width and with subfigures. The same concept works for tables, too.

\section{Progress in higher order relativistic chiral nuclear forces}

While non-relativistic chiral nuclear forces have been developed up to N$^4$LO (complete) and partial N$^5$LO, relativistic chiral nuclear forces still need to be extended to higher orders to check chiral convergence and improve the description of high-energy scattering data. Currently, we are constructing relativistic chiral NN forces up to N$^3$LO~\cite{Lu:2025ubc}. The Feynman diagrams to be calculated up to N$^3$LO can be specified by the following power-counting rules of ChEFT. The chiral order $\nu$ for a certain Feynman diagram with $L$ loops is counted as
\begin{equation}\label{eq:pcr1}  
   \nu = 4 L-2N_\pi-N_N+\sum_k kV_k 
\end{equation}
where $N_{\pi,N}$ are the numbers of $\pi$ and nucleon propagators, and $V_k$ is the number of $k$-th order vertices.

Thus, at N$^3$LO, the chiral force will involve:
\begin{enumerate}
\item Contact terms: 23 contact terms have been identified and included, encompassing scalar, vector, axial-vector, and tensor interactions~\cite{Xiao:2018jot}.

\item One-pion exchanges (1$\pi$): The one-pion exchange potential is expanded as $$V_{1\pi }=V_{1\pi }^{(0)}+V_{1\pi }^{(2)}+V_{1\pi }^{(3)}+V_{1\pi }^{(4)}+\cdots,$$ with contributions up to N$^3$LO fully considered. Isospin-breaking effects from the mass difference between charged and neutral pions in the one-pion exchange potential are explicitly incorporated. The contributions from higher-order OPE can be completely absorbed by taking $g_A=1.290$, and that of $f_\pi$, $m_\pi$, and $m_N$ can be taken care of by using their physical values. Such a treatment is equivalent to the non-relativistic treatments~\cite{Epelbaum:2002gb,Machleidt:2011zz} and it is valid up to N$^3$LO.

\item Two-pion exchanges (2$\pi$): The two-pion exchange potential is expressed as $$V_{2 \pi}=V_{2 \pi }^{(2)}+V_{2 \pi }^{(3)}+V_{2 \pi }^{(4)}+\cdots,$$ including both one-loop and two-loop contributions. The two-loop diagrams are the most challenging part, so we adopt the spectral representation method (dispersion relation/Mandelstam representation) instead of dimensional regularization—an approach more suitable for massive fermions in nuclear forces, which has also been successfully applied in recent developments of non-relativistic nuclear forces. 

\item Three-pion exchanges (3$\pi$): The three-pion exchange potential starts at N$^3$LO $$V_{3 \pi}=V_{3 \pi}^{(4)}+\cdots$$ and is found to be negligible, thus completely absorbed in the higher-order corrections, at least in non-relativistic studies. We temporarily adopt such a treatment.  

\item Isospin breaking effects: All isospin breaking effects—including charge-independence breaking (CIB), charge-symmetry breaking (CSB), and the Coulomb force for pp interactions—are included to ensure the accuracy of the force in describing isospin-asymmetric nuclear systems. Two additional charge-dependent contact operators are also incorporated to further improve precision.
\end{enumerate}

Given the extreme complexity of the dimensional regularization method for multi-loop calculations, we adopted the spectral-function regularization method for all loop diagrams at NLO, N$^2$LO, and N$^3$LO in our calculations. A Gaussian-type form factor was used to suppress the high-momentum components of the exchanged pions, and the cutoff was varied from $0.5$ to $0.8$ GeV as a rough estimate of the theoretical uncertainties. 

\begin{figure}[H]
%\isPreprints{}{% This command is only used for ``preprints''.
\begin{adjustwidth}{-\extralength}{0cm}
\centering
\subfloat[\centering]{\includegraphics[width=5.0cm]{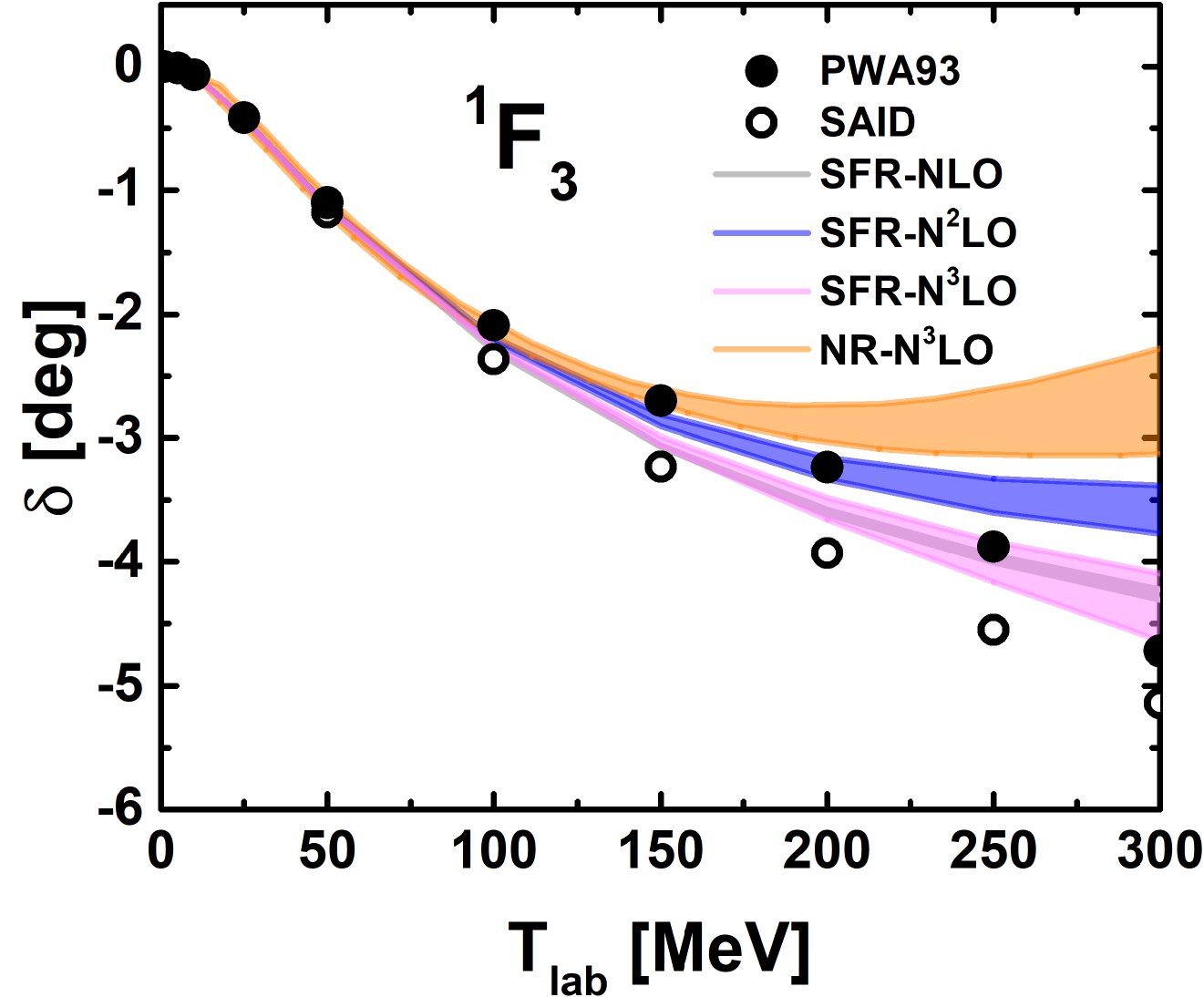}}
%\hfill
\subfloat[\centering]{\includegraphics[width=5.0cm]{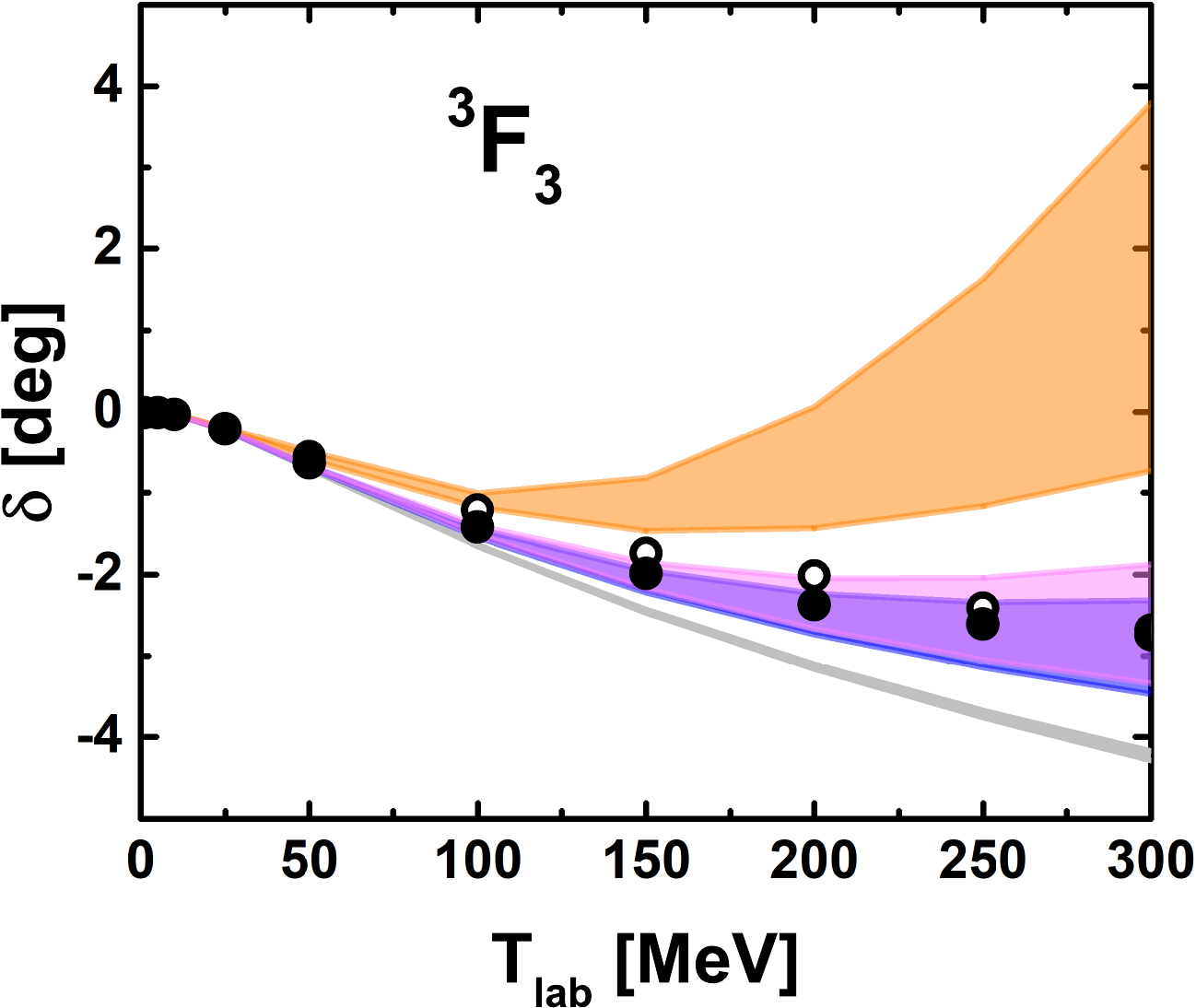}}
\subfloat[\centering]{\includegraphics[width=5.0cm]{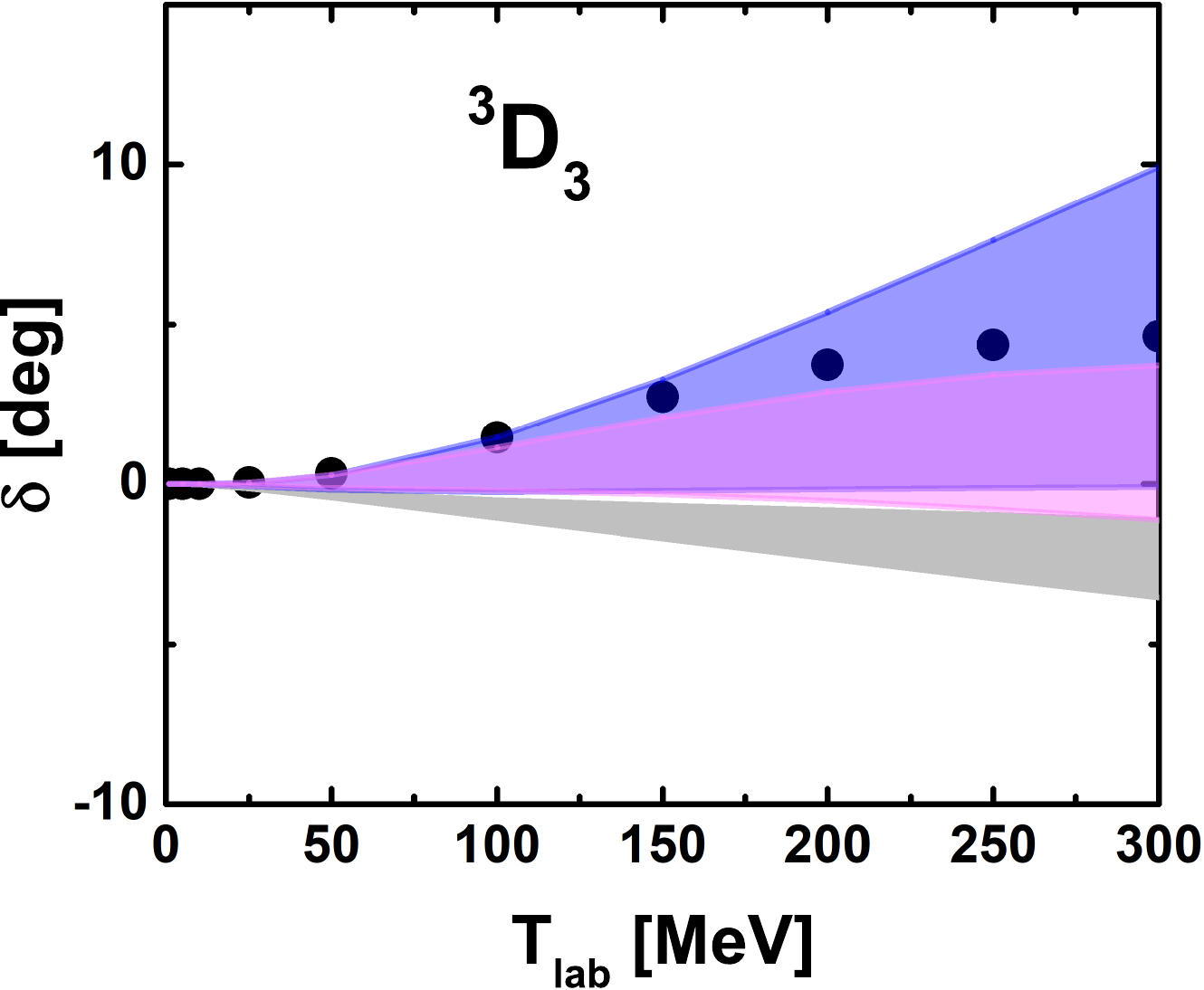}}\\
\subfloat[\centering]{\includegraphics[width=5.0cm]{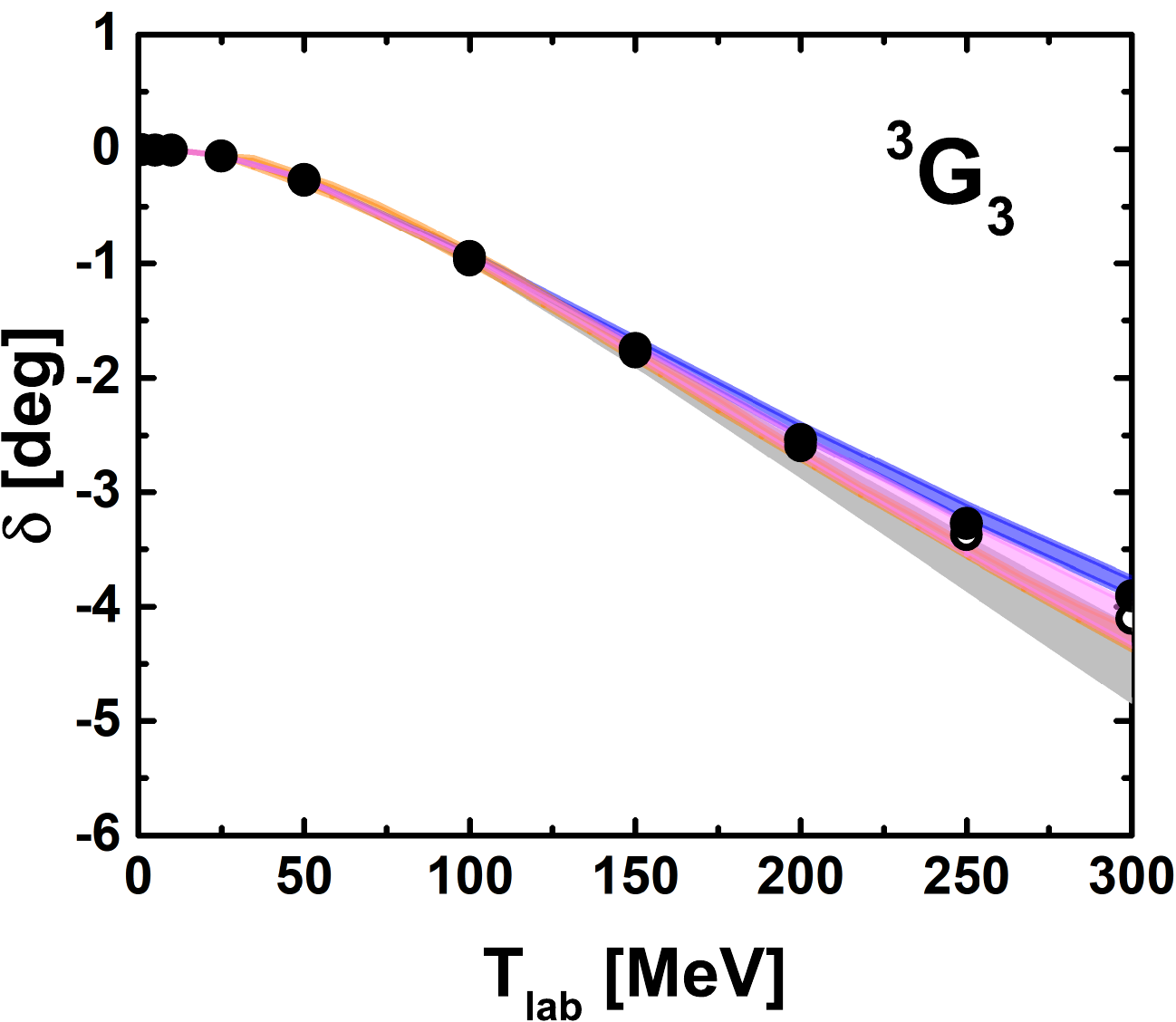}}
%\hfill
\subfloat[\centering]{\includegraphics[width=5.0cm]{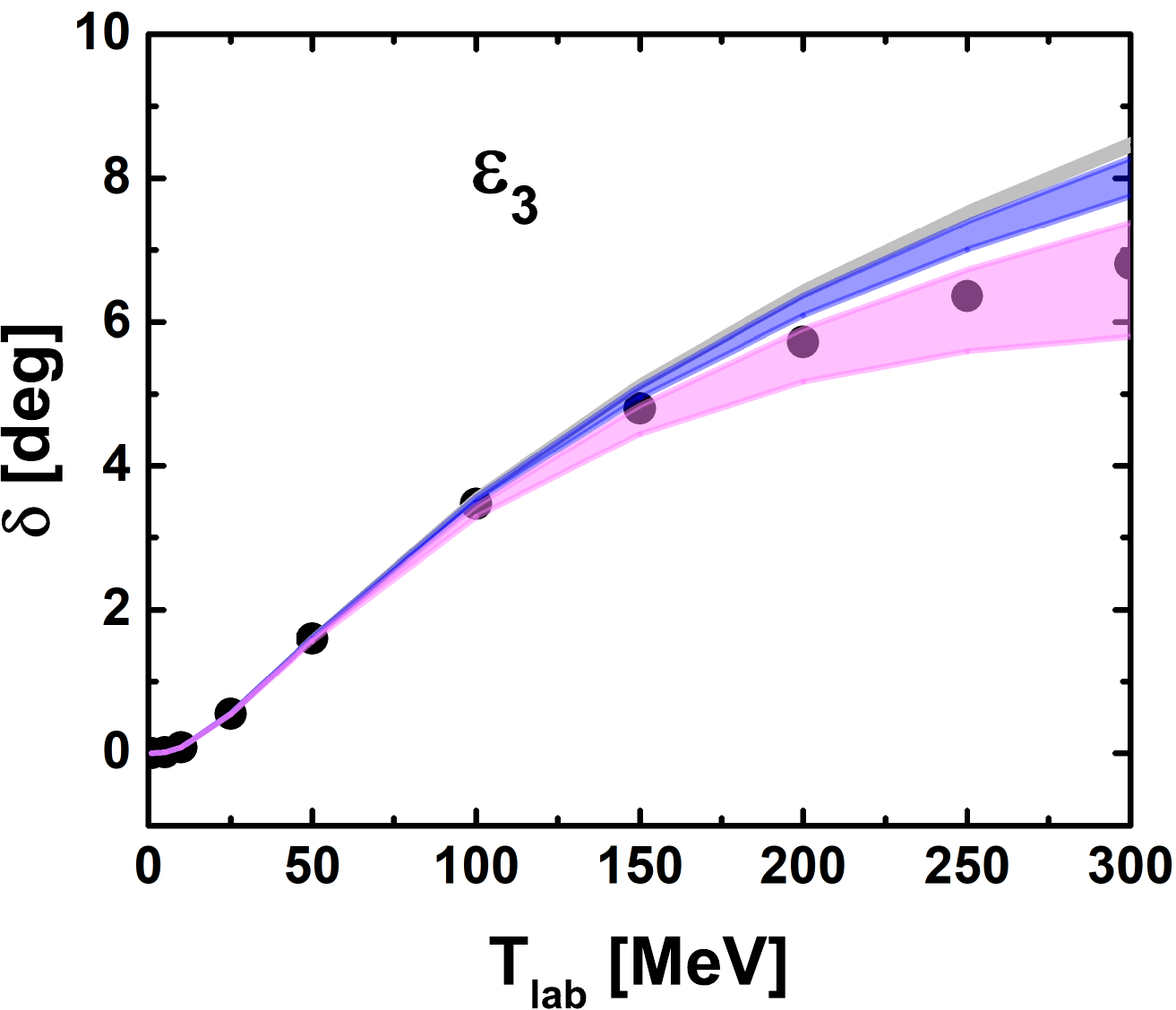}}\\
%\isPreprints{}{% This command is only used for ``preprints''.
\end{adjustwidth}
\caption{Phase shifts and mixing angle for the $J=3$ partial waves. The gray, blue, and magenta bands are the results from the relativistic TPE $NN$ interactions at NLO, N$^2$LO, and N$^3$LO, respectively, with a cutoff in the range $[0.5, 0.8]$ GeV. For comparison, the N$^3$LO non-relativistic results are shown with orange bands (the two $F$-waves from Ref.~\cite{Entem:2014msa} and the $G$-waves from Ref.~\cite{Entem:2015xwa}). The solid and open dots represent the data from the Nijmegen multi-energy neutron-proton (n-p) phase shift analysis~\cite{Stoks:1993tb} and the VPI/GWU single-energy n-p analysis SM99~\cite{Arndt:1994br}, respectively. Taken from Ref.~\cite{Lu:2025ubc}. \label{fig3}}
\end{figure}

\begin{figure}[H]
%\isPreprints{}{% This command is only used for ``preprints''.
\begin{adjustwidth}{-\extralength}{0cm}
\centering
\subfloat[\centering]{\includegraphics[width=5.0cm]{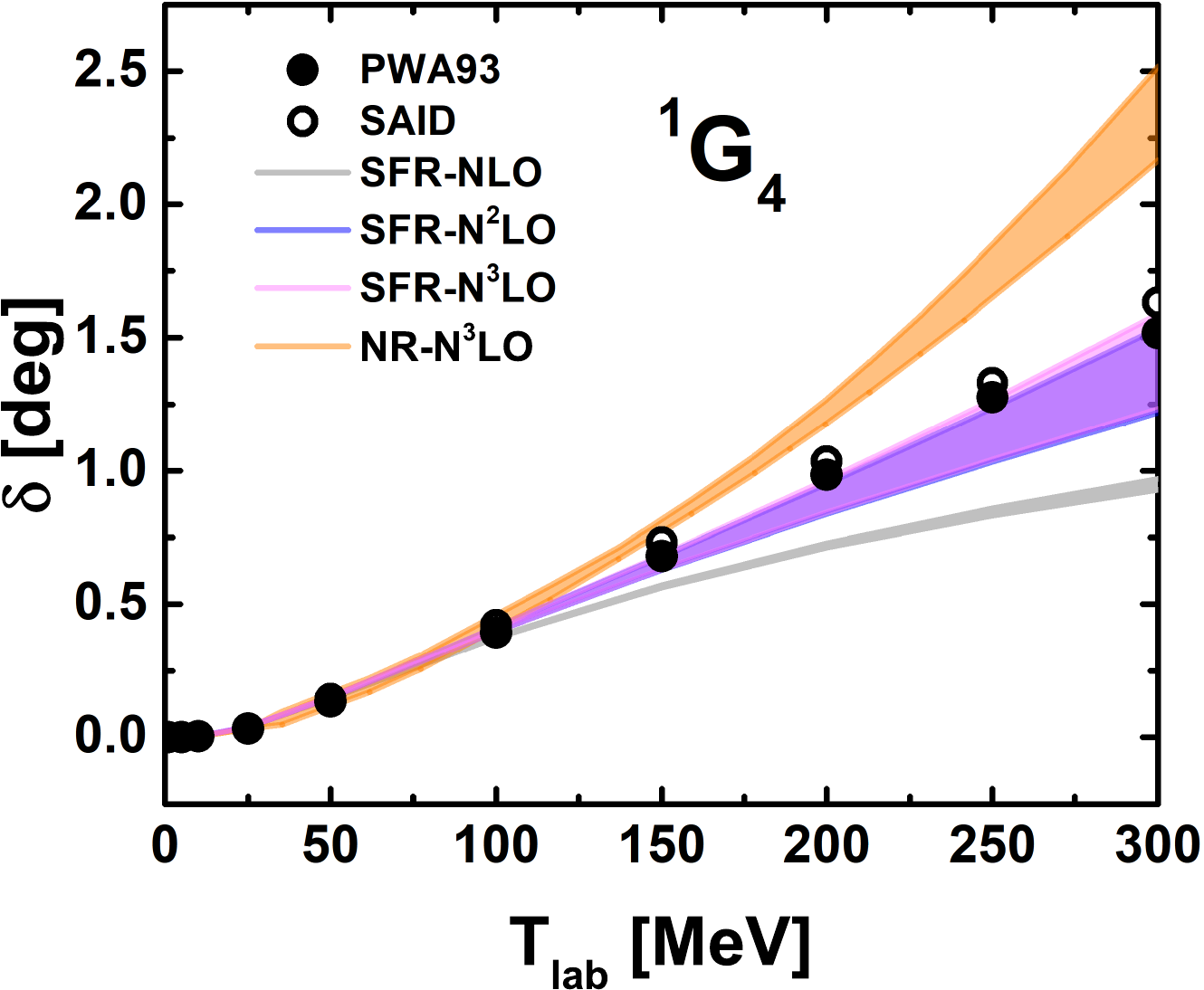}}
%\hfill
\subfloat[\centering]{\includegraphics[width=5.0cm]{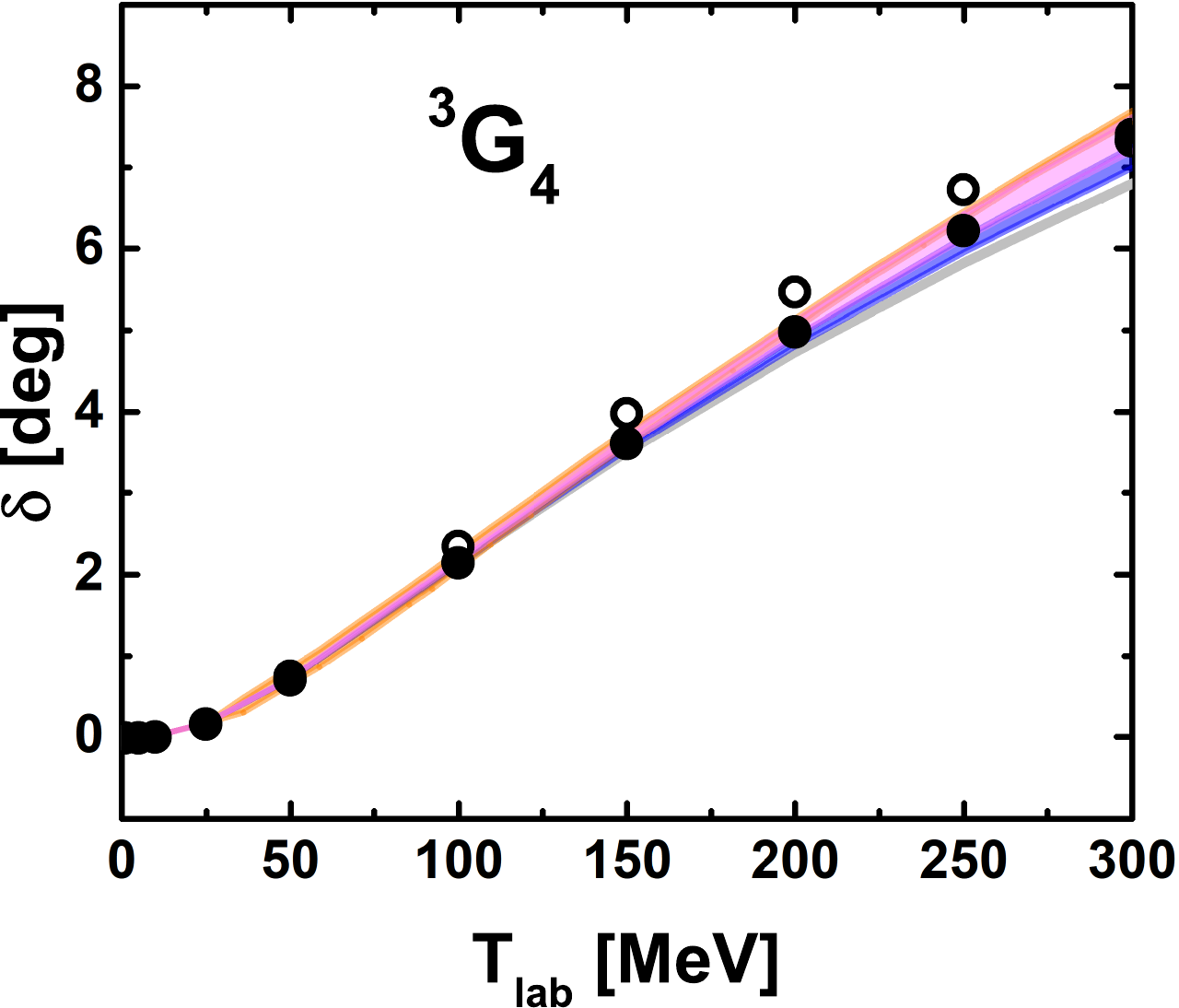}}
\subfloat[\centering]{\includegraphics[width=5.0cm]{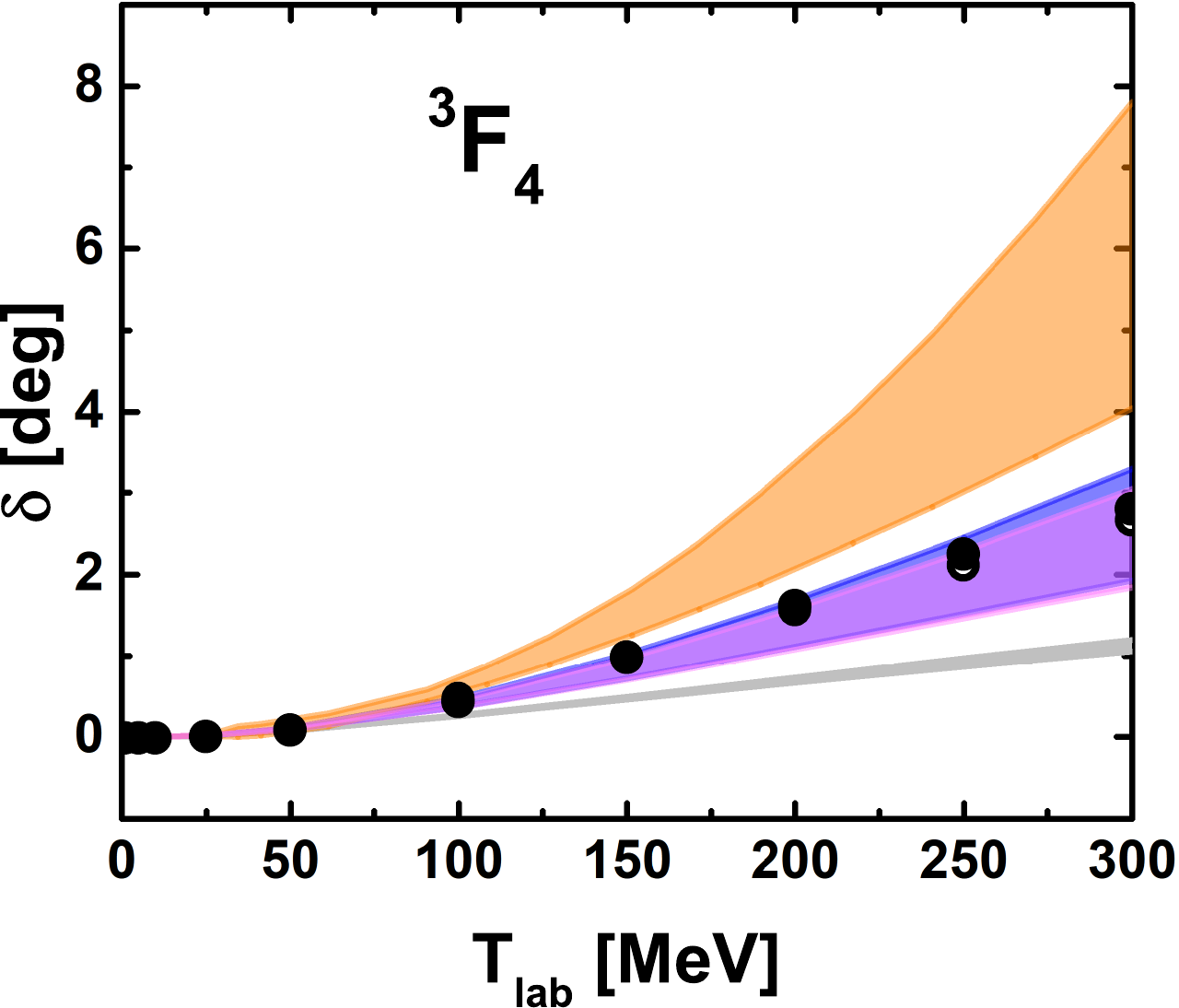}}\\
\subfloat[\centering]{\includegraphics[width=5.0cm]{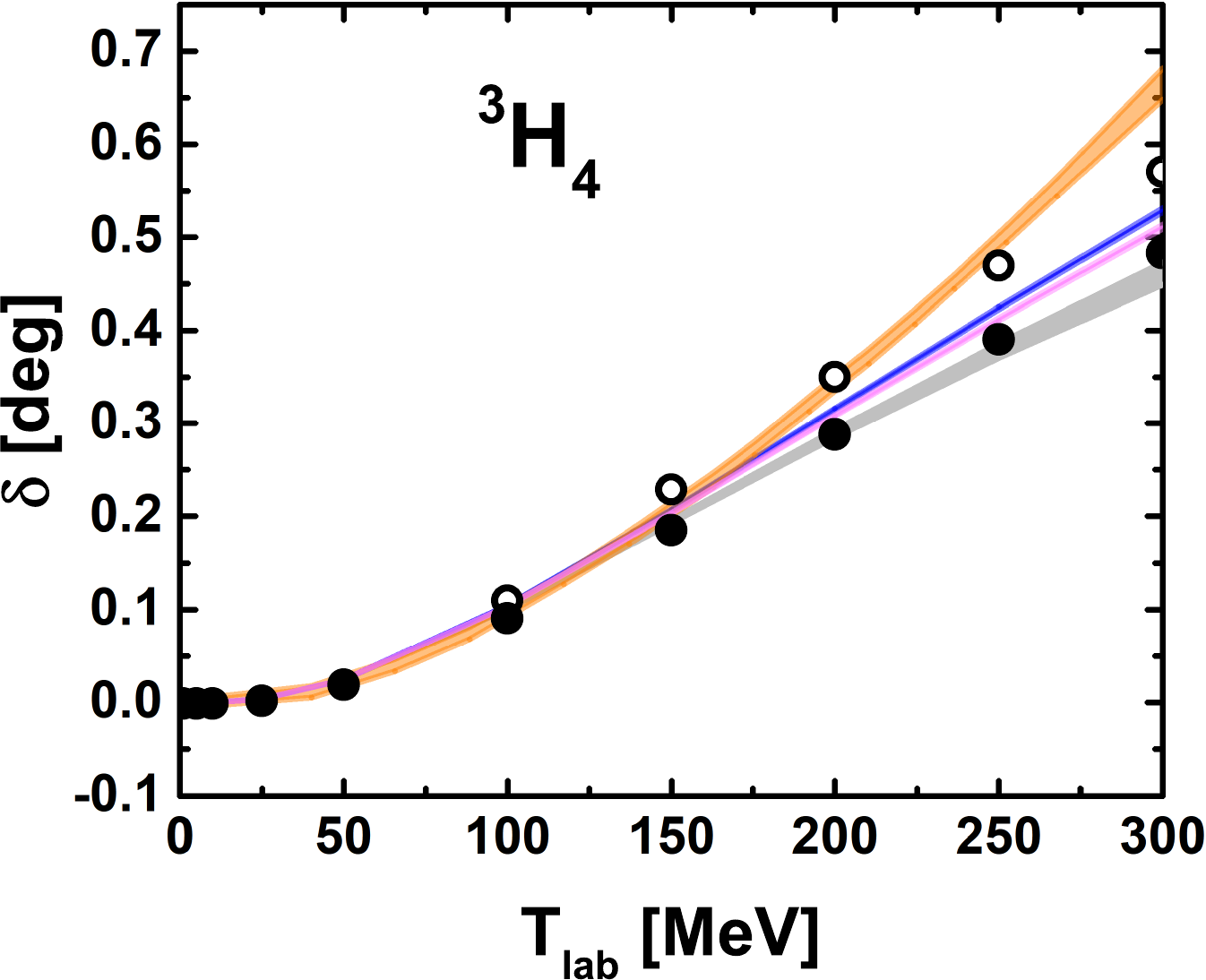}}
%\hfill
\subfloat[\centering]{\includegraphics[width=5.0cm]{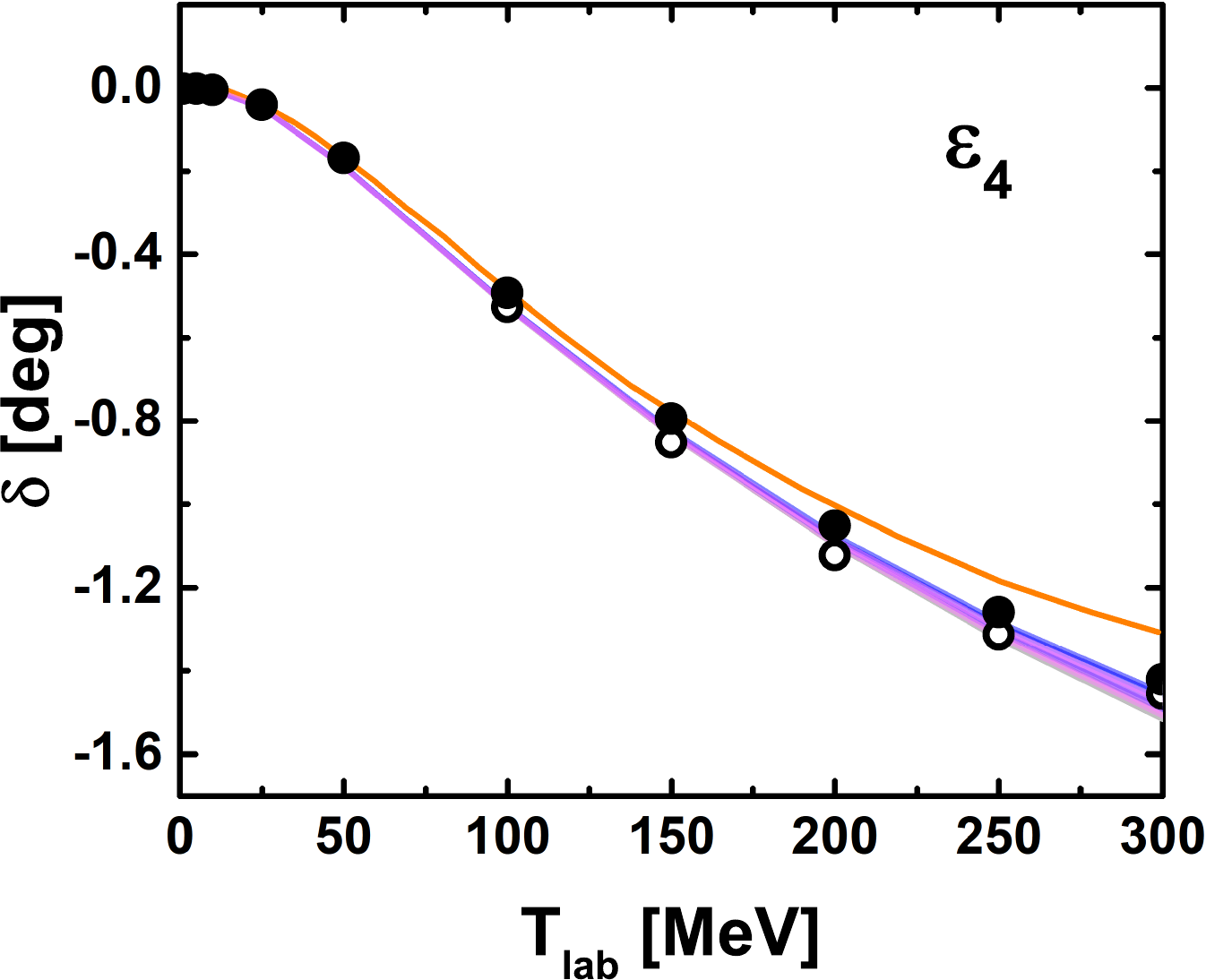}}\\
%\isPreprints{}{% This command is only used for ``preprints''.
\end{adjustwidth}
\caption{Same as Figure~\ref{fig3} but for the $J=4$ partial waves. The solid and open dots are the data from the Nijmegen multi-energy n-p phase shift analysis~\cite{Stoks:1993tb} and the GWU n-p analysis SP07~\cite{Arndt:2007qn}. Taken from Ref.~\cite{Lu:2025ubc}.\label{fig4}}
\end{figure}

\begin{figure}[H]
%\isPreprints{}{% This command is only used for ``preprints''.
\begin{adjustwidth}{-\extralength}{0cm}
\centering
\subfloat[\centering]{\includegraphics[width=5.0cm]{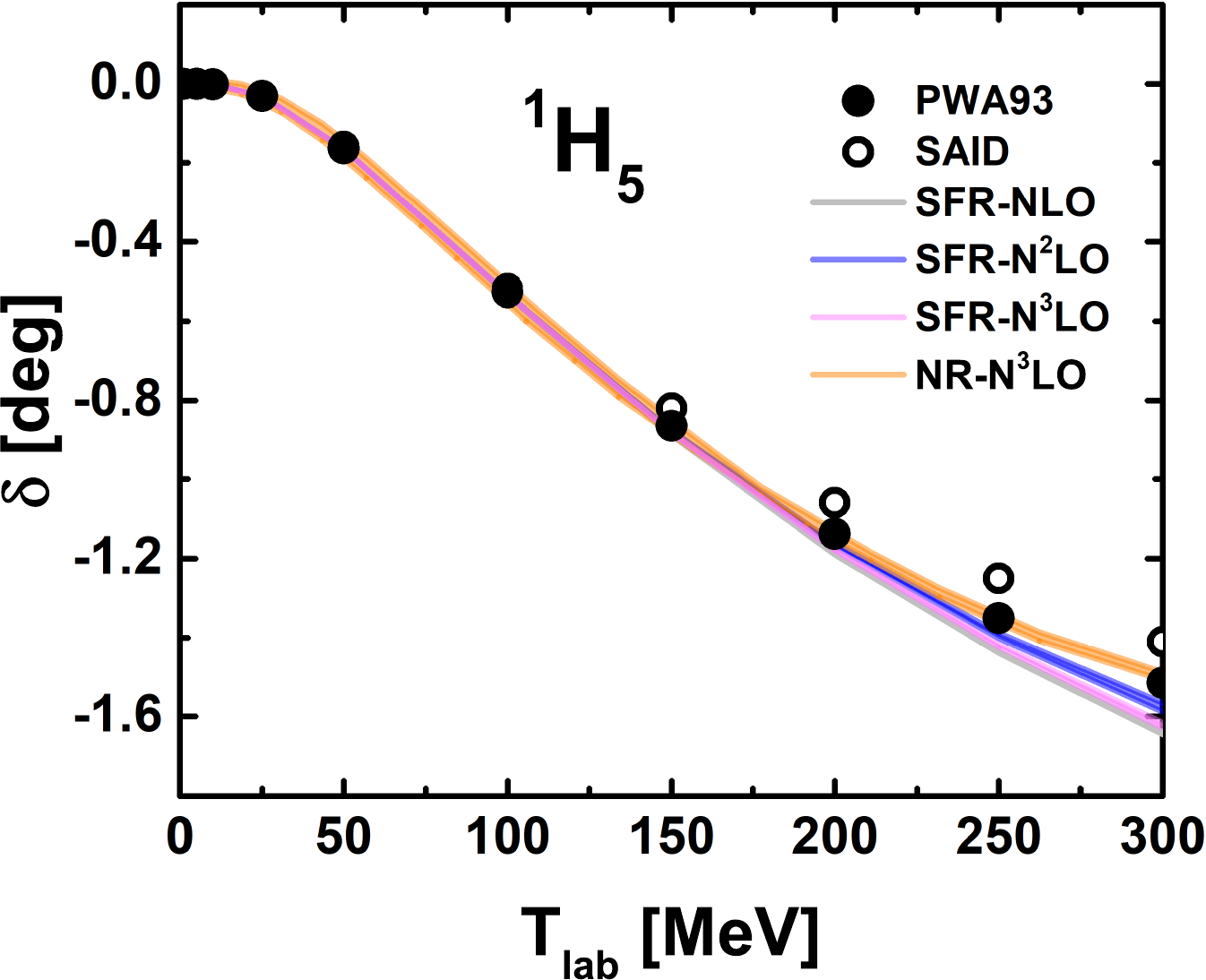}}
%\hfill
\subfloat[\centering]{\includegraphics[width=5.0cm]{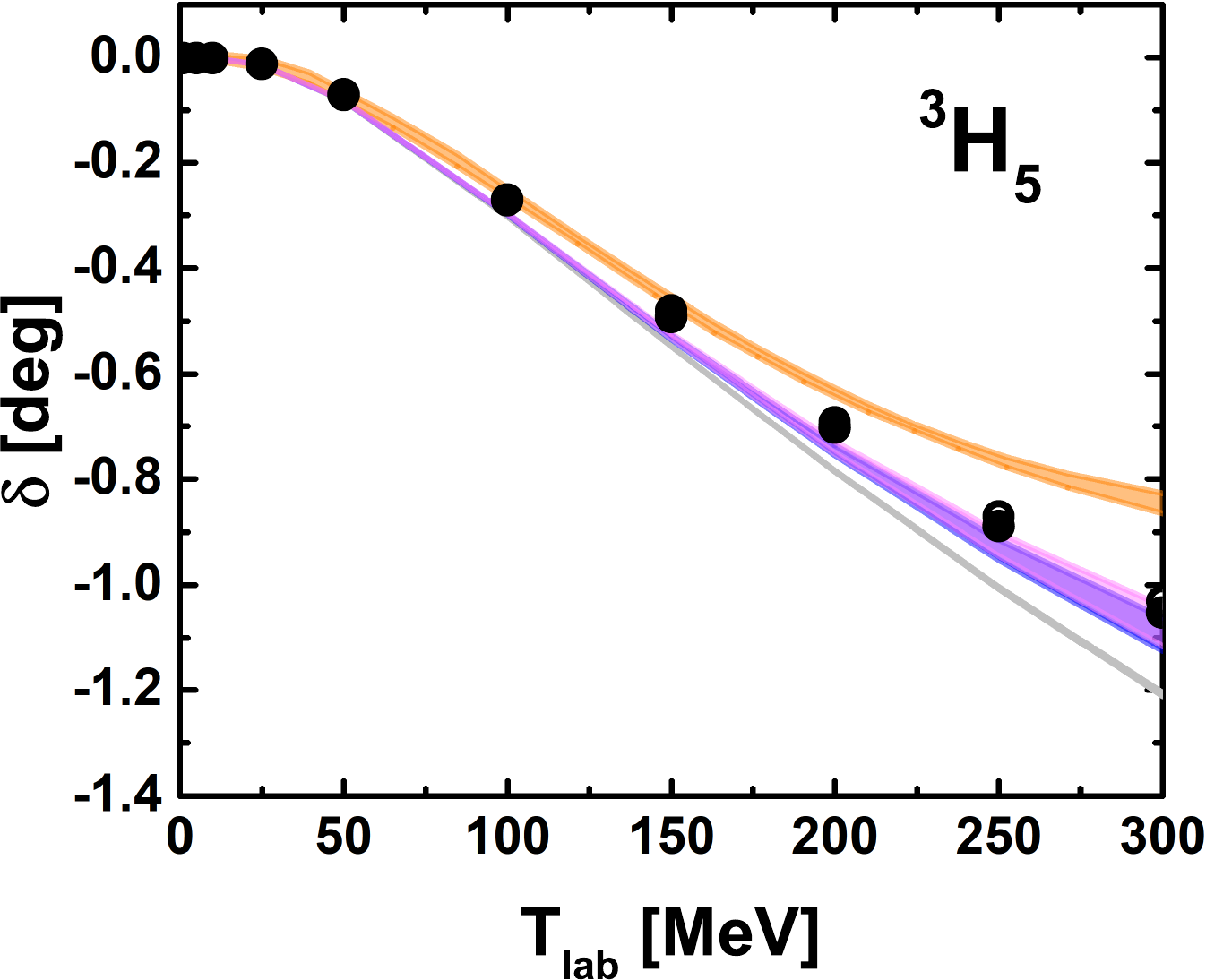}}
\subfloat[\centering]{\includegraphics[width=5.0cm]{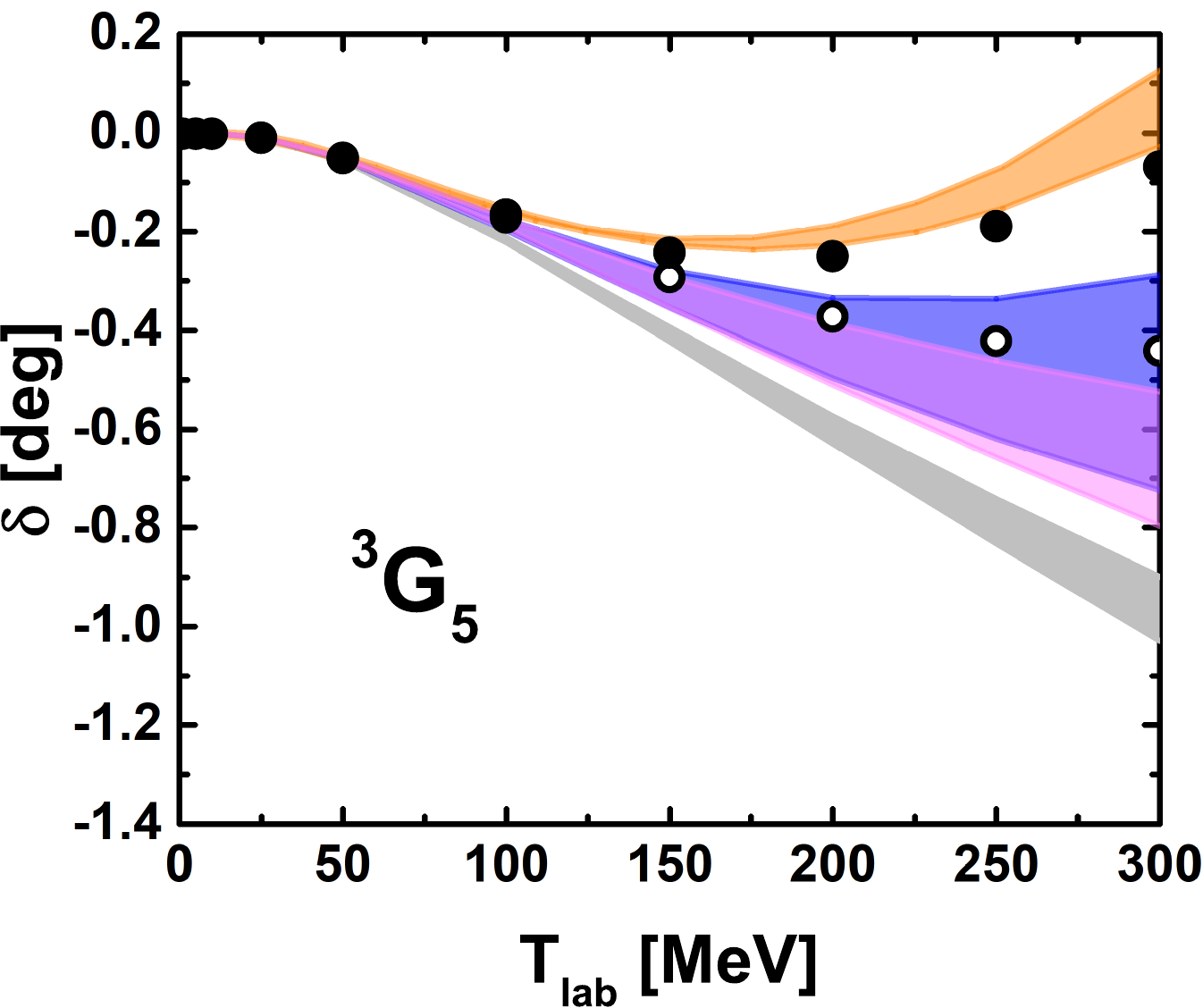}}\\
\subfloat[\centering]{\includegraphics[width=5.0cm]{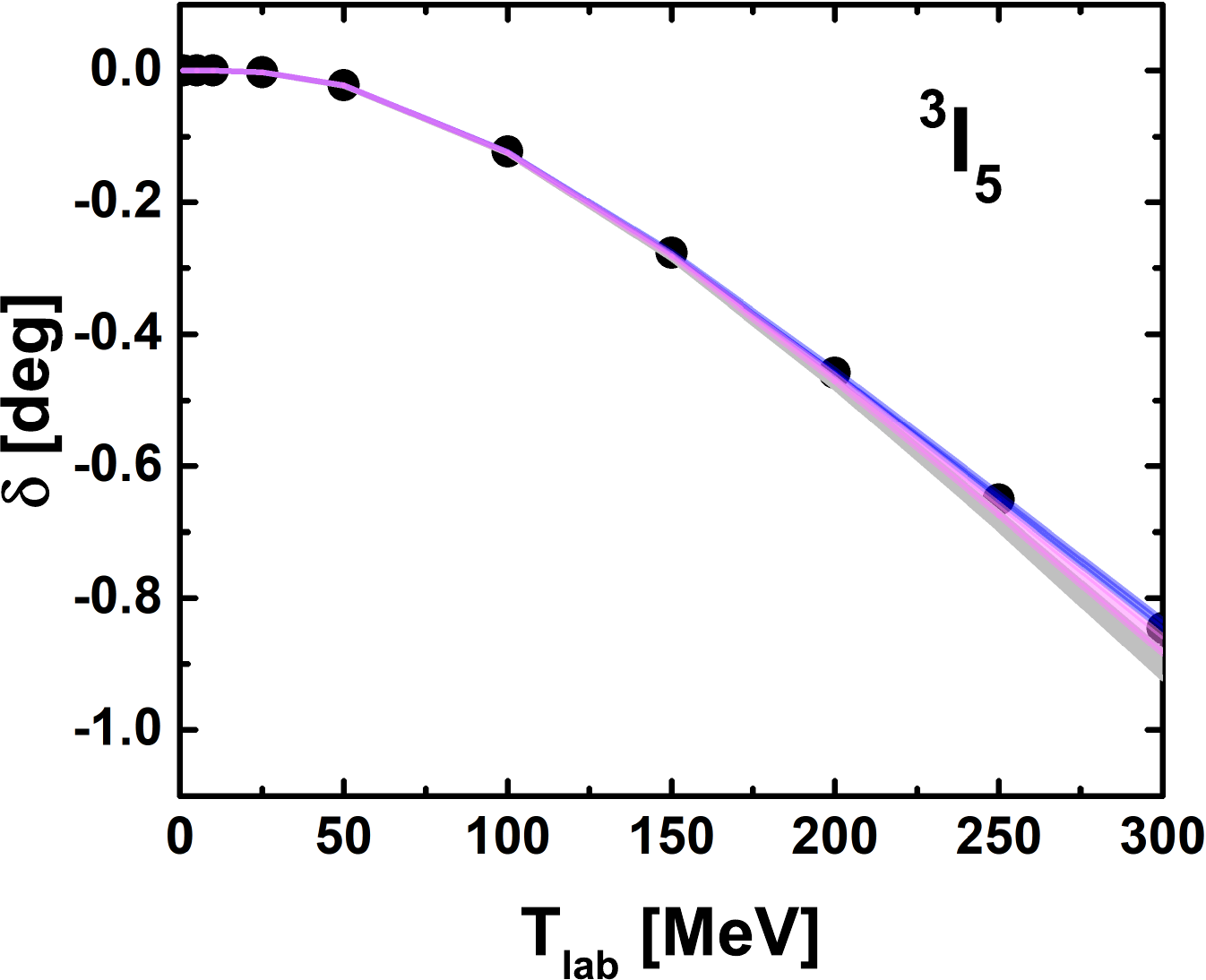}}
%\hfill
\subfloat[\centering]{\includegraphics[width=5.0cm]{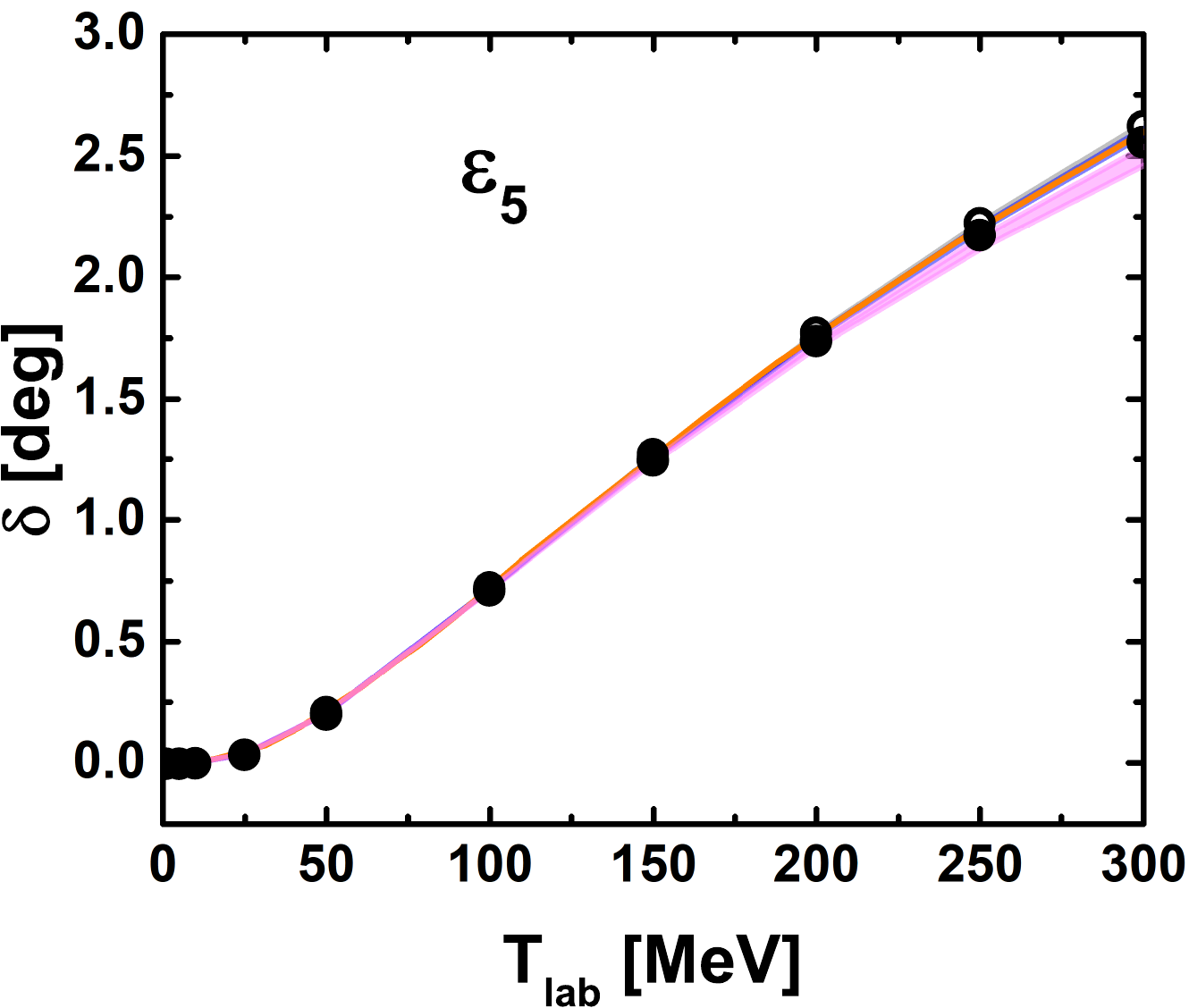}}\\
%\isPreprints{}{% This command is only used for ``preprints''.
\end{adjustwidth}
\caption{Same as Figure~\ref{fig4} but for the $J=5$ partial waves. Taken from Ref.~\cite{Lu:2025ubc}.\label{fig5}}
\end{figure}

In Figs.~\ref{fig3},~\ref{fig4}, and~\ref{fig5}, we present perturbative results at NLO, NNLO, and N$^3$LO and compare them with the non-relativistic results. In these peripheral partial waves of interest, the phase shifts are in very good agreement with the PWA93 and SAID data. Notably, good agreement is already achieved up to the NNLO order for most partial waves, while noticeable corrections from the fourth chiral order (N$^3$LO) only appear in very few partial waves—specifically those with smaller orbital angular momentum $L$ in the higher energy region, such as the $^1F_3$ partial wave. Unlike non-relativistic calculations, where contributions from the fifth chiral order (N$^4$LO) are still required to provide repulsive interactions that compensate for the excessive attraction from lower chiral orders~\cite{Entem:2014msa}, this pattern points to a faster convergence of the relativistic chiral nuclear force. 

% Example of a figure that spans the whole page width and with subfigures. The same concept works for tables, too.

% Example of a figure that spans the whole page width and with subfigures. The same concept works for tables, too.

\subsection{Progress in the development of relativistic three-body scattering equation}

In non-relativistic studies, 3N forces are essential for reproducing triton binding energies and nuclear matter saturation. Theoretically, they naturally emerge in EFTs when integrating out high-momentum modes. In relativistic chiral theories, constructing 3N forces is more complex but necessary to achieve self-consistent descriptions of nuclear systems. On the other hand, recent relativistic studies have shown that nuclear matter saturation and reasonable descriptions of medium-mass nuclei can be achieved with LO relativistic chiral forces without introducing 3N forces~\cite{Shen:2025iue}, highlighting the efficiency of relativistic frameworks.
Based on the Faddeev equation, we have established a theoretical framework for three-nucleon (3N) scattering that is suitable for calculations with relativistic two-body chiral nuclear forces and naturally extends to three-nucleon interactions, a key focus of recent progress in relativistic chiral nuclear force research. By generalizing the non-relativistic Faddeev equation to the relativistic case, we derive the relativistic three-body scattering equation, which has a form similar to the Bethe-Salpeter equation (BSE) in relativistic two-body scattering, thereby ensuring Lorentz invariance and self-consistency. The relativistic three-body scattering equation can be expressed in a form that is formally identical to the Faddeev equation:

$$\left(\begin{array}{c}T^{(i)} \\ T^{(j)} \\ T^{(k)}\end{array}\right)=\left(\begin{array}{c}t_{i} \\ t_{j} \\ t_{k}\end{array}\right)+\left(\begin{array}{ccc}0 & t_{i} & t_{i} \\ t_{j} & 0 & t_{j} \\ t_{k} & t_{k} & 0\end{array}\right) G_{0}\left(\begin{array}{c}T^{(i)} \\ T^{(j)} \\ T^{(k)}\end{array}\right)$$
where \(t_i = v_i + v_i G_0 t_i\) is the relativistic two-body t-matrix, \(G_0\) is the relativistic propagator of three nucleons, and \(v_i\) includes all irreducible connected Feynman diagrams (two-body interactions). For the study of neutron–deuteron scattering, the aforementioned system of integral equations provides the starting point and can be written in the following compact form, which is similar to the Alt-Grassberger-Sandhas (AGS) equation
\begin{align}
    U=Pv\frac{E_q}{m}(2\pi)^3+PtG_0U,
    \label{AGS equation}
\end{align}
where we suppress the explicit notation of the initial and final states, which consist of a wave function generated by the deuteron vertex and a free nucleon, $v$ is the $NN$ interaction, $t$ is the two-body $t$-matrix defined by the BSE, $P=P_{12}P_{23}+P_{13}P_{23}$ is the permutation operator, which determines the overlap between different Jacobi channels, and
\begin{align}
    G_0=\frac{1}{(2\pi)^3}\frac{m^2}{E_p^2}\frac{1}{\sqrt{(\sqrt{s}-E_q)^2-q^2}-2E_p+i\epsilon},
\end{align}
is the free three-particle propagator, where $m$ is the mass of the nucleon, $\sqrt{s}$ is the energy of the three-body system, $p$ and $q$ denote the standard Jacobi momenta, $E_q=\sqrt{q^2+m^2}$, and $E_p=\sqrt{p^2+m^2}$. Note that we have considered protons and neutrons as identical particles with isospin $1/2$.

\begin{figure}[H]
%\isPreprints{\centering}{} % Only used for preprints
\includegraphics[width=14.0 cm]{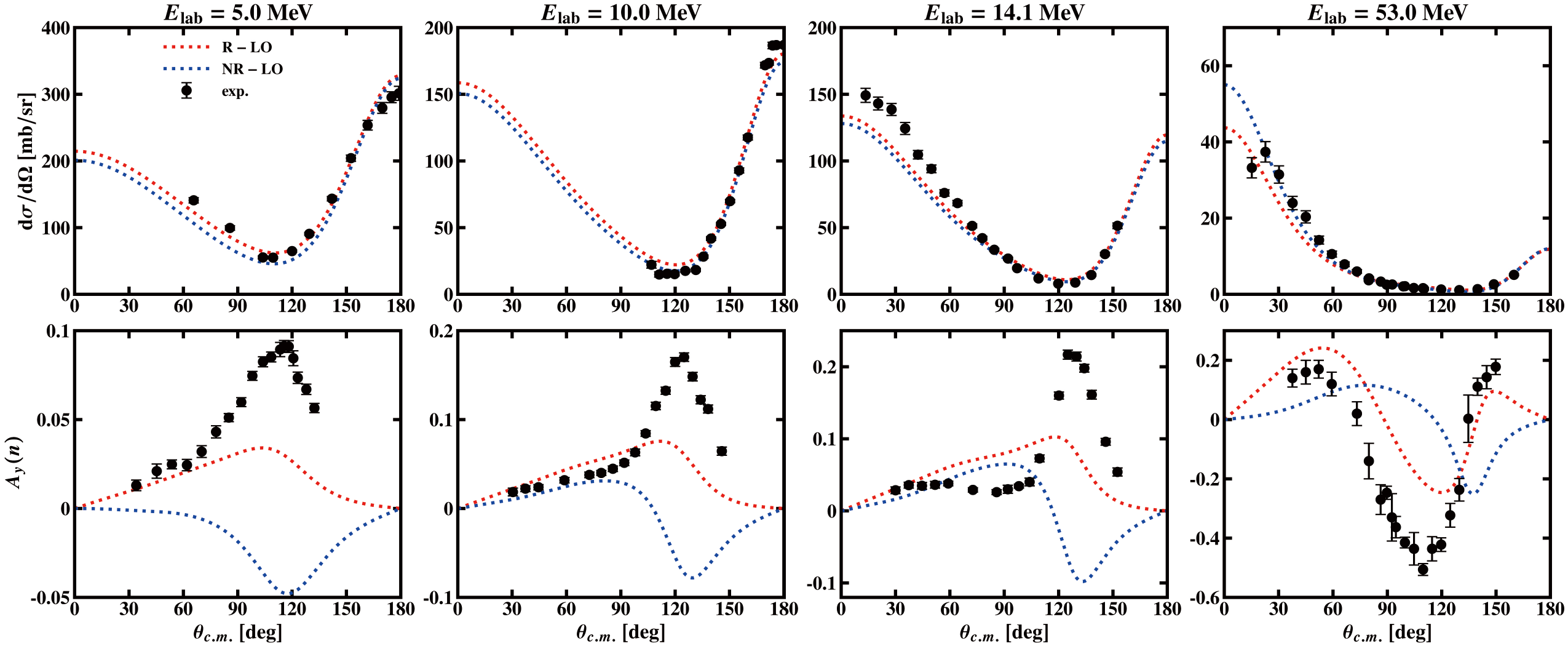}
\caption{~(Color online)~The differential cross section  and the neutron analyzing powers $A_y$ for elastic $nd$ scattering with the incident neutron energy $E_{lab} = 5.0, 10.0, 14.1, 53.0$ MeV. The red and blue dotted lines denote the results obtained from the relativistic LO chiral force and its non-relativistic LO counterpart. The black dots with error bars are experimental data from Refs.~\cite{Otuka:2014wzu,Berick:1968zz,Romero:1982zz,Watson:1982zz}.\label{fig:ndLO}}
\end{figure}

Preliminary studies of the differential cross section and the neutron analyzing powers $A_y$ for $nd$ scattering in the relativistic framework (at \(E_{lab} = 5.0, 10.0, 14.1, 53.0\) MeV) are shown in Figure~\ref{fig:ndLO}, yield promising results that may help address the $A_y$ puzzle and improve the description of polarization observables~\cite{Zhai:2026aaa}.

\section{Recent applications in nuclear and hypernuclear systems}

\subsection{Symmetric nuclear matter}

We have applied relativistic chiral nuclear forces within the Relativistic-Brueckner-Hartree-Fock (RBHF) theory to study symmetric nuclear matter and compared the results with those obtained using the non-relativistic Brueckner-Hartree-Fock (BHF) theory. This research program, initiated in Ref.~\cite{Zou:2023quo} with LO and extended to next-to-leading order (NLO) in Ref.~\cite{Zou:2025dao}, represents the first covariant chiral force studies in the RBHF framework, bridging the gap between relativistic and non-relativistic \textit{ab initio} approaches.

\begin{figure}[H]
\includegraphics[width=7.5 cm]{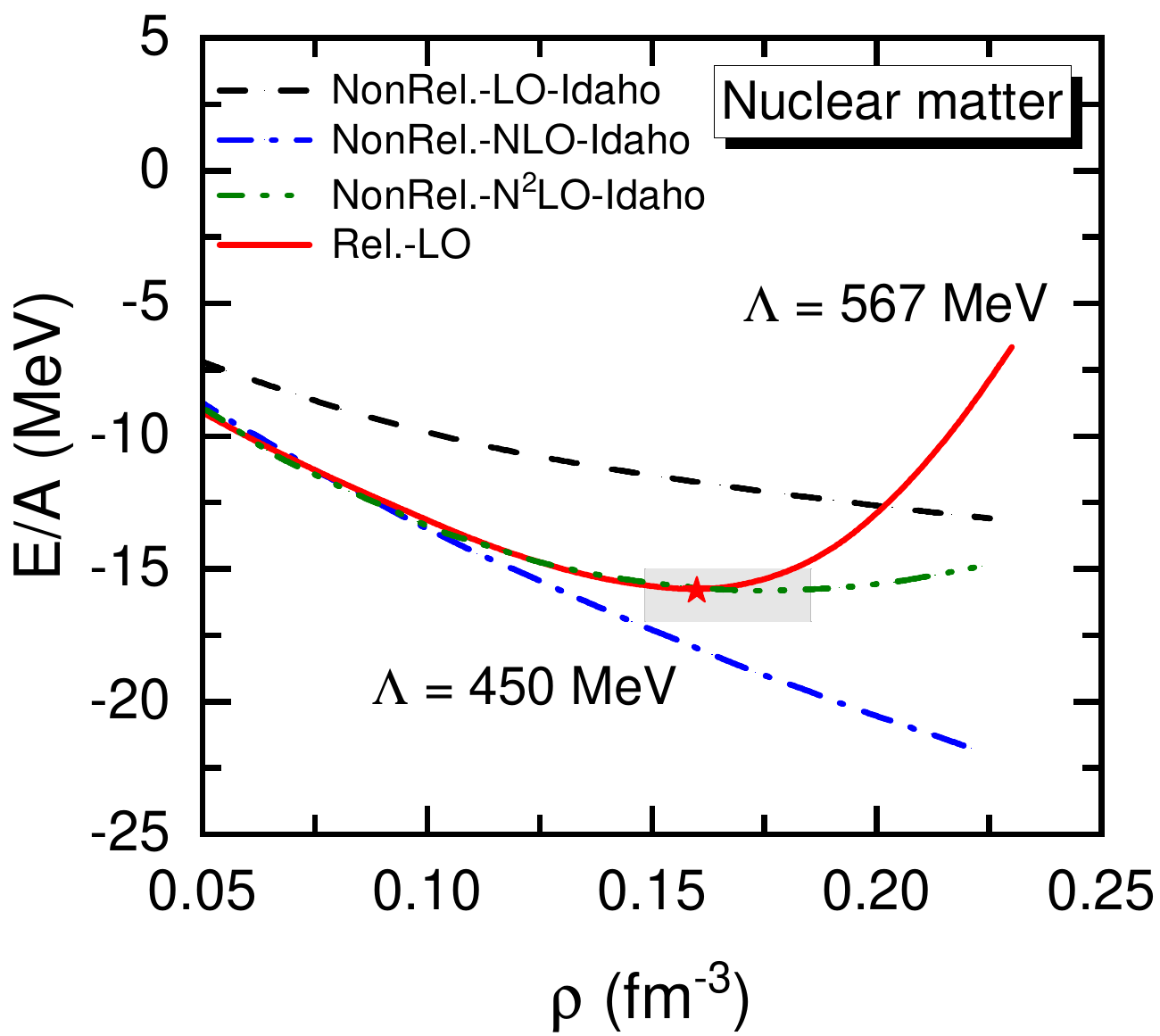}
\caption{Energy per nucleon in symmetric nuclear matter. The red solid line shows the LO RBHF result from Ref.~\cite{Zou:2023quo} ($\Lambda=567$ MeV), compared with non-relativistic BHF results at LO, NLO, and N$^2$LO~\cite{Sammarruca:2021bpn}. The shaded area indicates the empirical saturation region~\cite{Beth71,Spru72}. Taken from Ref.~\cite{Zou:2023quo}.\label{fig7}}
\end{figure}

In non-relativistic BHF theory, nuclear matter saturation cannot be achieved up to NLO; only at N$^2$LO, with three-nucleon forces (3NFs) included, can saturation be realized. This highlights the persistent challenges faced by non-relativistic frameworks in describing this fundamental nuclear property. By contrast, in the RBHF theory, using only four LECs at LO already achieves saturation without explicit 3NFs, as shown in Figure~\ref{fig7}. While the LO RBHF result saturates near the empirical region, the non-relativistic LO and NLO results do not, indicating that nuclear matter saturation can be understood as a relativistic effect~\cite{Zou:2023quo}.

\begin{figure}[h]
\includegraphics[width=0.45\textwidth]{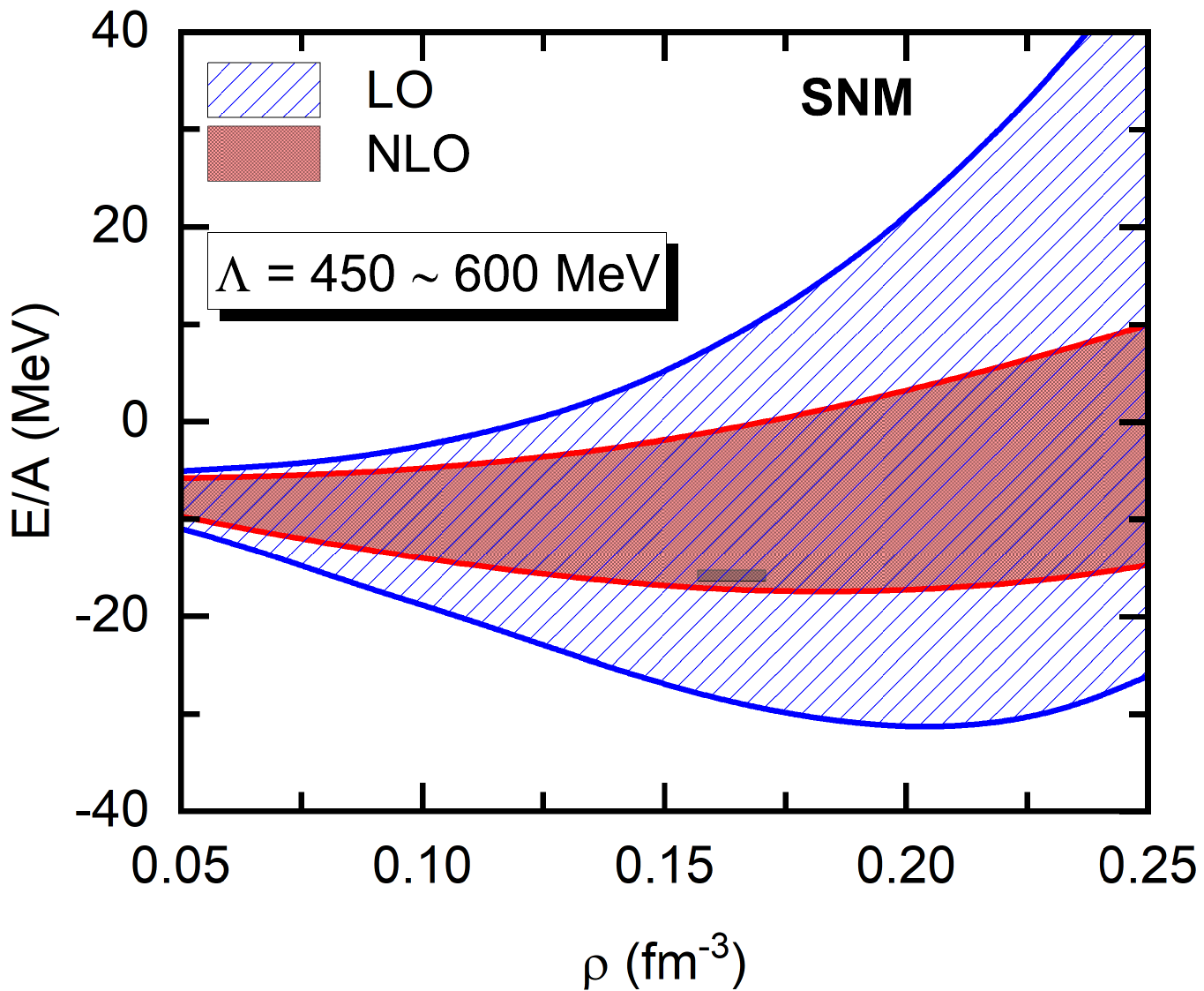}
\caption{Energy per nucleon in symmetric nuclear matter obtained with covariant chiral nuclear forces: LO (blue band) from Ref.~\cite{Zou:2023quo} and NLO (red band) from Ref.~\cite{Zou:2025dao}, with cutoff variation $\Lambda=450\text{-}600$ MeV. The reduced bandwidth at NLO demonstrates systematic convergence. Taken from Ref.~\cite{Zou:2025dao}.\label{fig6}}
\end{figure}

The NLO extension demonstrates systematic order-by-order convergence. Figure~\ref{fig6} compares LO and NLO results with cutoff variations $\Lambda=450\text{-}600$ MeV. The NLO band lies completely within the LO band as expected from chiral effective field theory, with the cutoff uncertainty reduced by approximately half (from $\sim$30 MeV to $\sim$16 MeV at saturation density). A key methodological advance in the NLO study~\cite{Zou:2025dao} was enforcing the naturalness of the 17 LECs to ensure numerical stability in nuclear matter calculations. At $\Lambda=590$ MeV, the NLO results yield excellent agreement with empirical values: saturation energy $-16.05$ MeV (empirical: $-16\pm1$ MeV), saturation density $0.167$ fm$^{-3}$ (empirical: $0.16\pm0.01$ fm$^{-3}$), and incompressibility $K_\infty=270$ MeV (empirical: $240\pm20$ MeV). The NLO equation of state becomes softer above saturation density than at LO, consistent with improved descriptions of the P-wave phase shift.

These results establish that \textit{ab initio} calculations of nuclear matter using covariant chiral nuclear forces display systematic order-by-order convergence, with NLO already providing quantitative saturation properties without explicit three-nucleon forces. The success of this approach reinforces the view that nuclear matter saturation can be treated as a relativistic effect, while the reduced cutoff dependence at NLO demonstrates the power of chiral effective field theory to control theoretical uncertainties. For pure neutron matter, the LO and NLO equations of state show substantially smaller cutoff uncertainties than symmetric nuclear matter, making these results particularly relevant for neutron star applications~\cite{Zou:2025dao}.

\subsection{Finite nuclei}
We have also applied relativistic chiral nuclear forces within the RBHF framework to study finite nuclei (e.g., $^{40}$Ca, $^{48}$Ca, $^{56}$Ni, $^{62}$Ni, $^{90}$Zr, $^{120}$Sn)~\cite{Shen:2025iue}. The results show that relativistic chiral nuclear forces can accurately describe the ground-state properties of finite nuclei, including charge density distributions, binding energies, and shell structures. The charge-density distributions calculated using relativistic chiral forces are in good agreement with experimental data, particularly for medium-mass and heavy nuclei. For example, the charge density of $^{120}$Sn is in excellent agreement with experimental results, demonstrating the efficiency of the relativistic framework.

\begin{figure}[H]
%\isPreprints{\centering}{} % Only used for preprints
\includegraphics[width=7.0 cm]{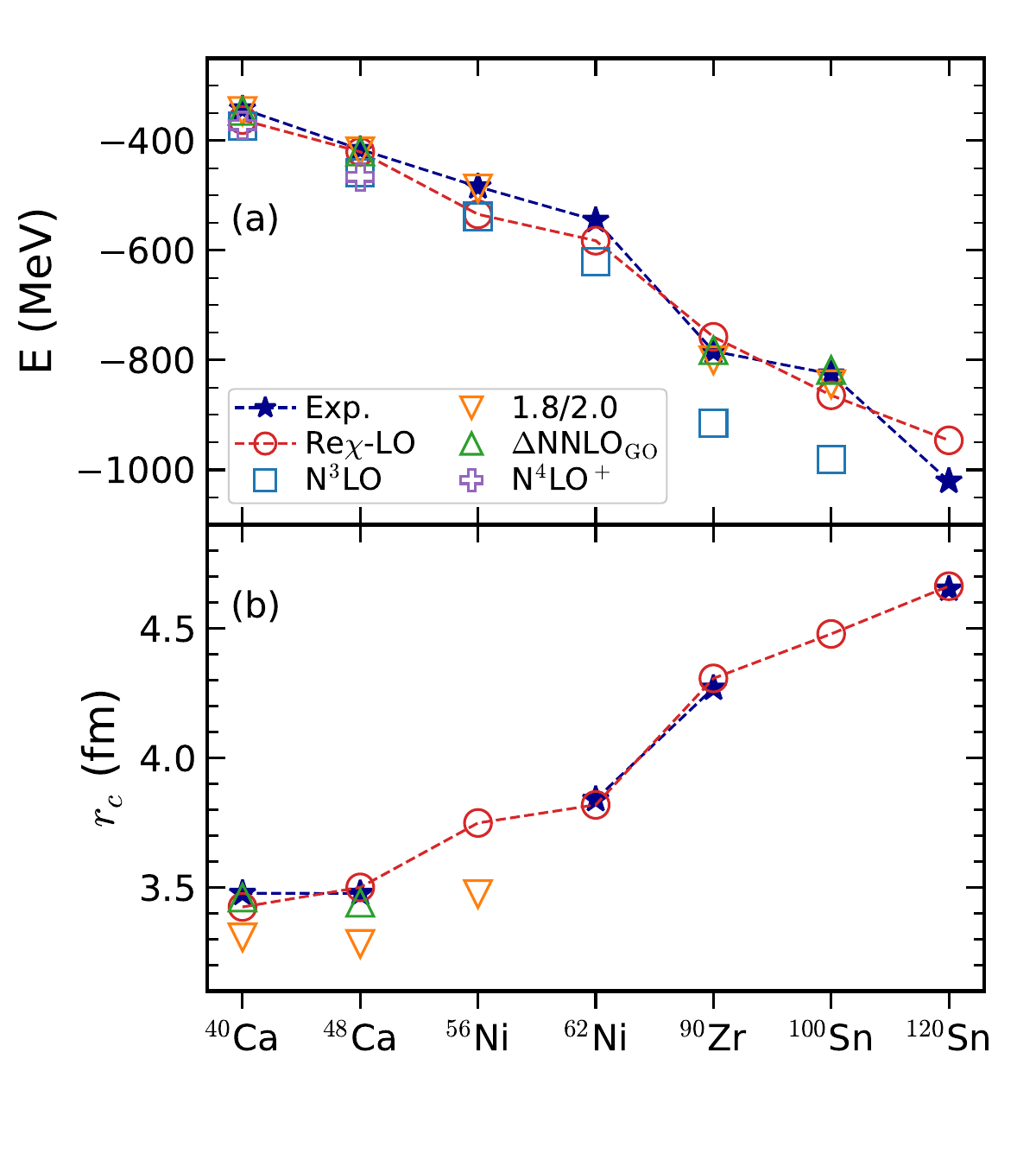}
\caption{Energies (a) and charge radii (b) of nuclei ($A = 40$ to $120$) from RBHF using the LO relativistic chiral force, compared to the nonrelativistic chiral forces~\cite{Binder:2013xaa,Morris:2017vxi,Simonis:2017dny,Jiang:2020the,LENPIC:2022cyu} and experimental data~\cite{Wang:2021xhn,Angeli:2013epw}. Taken from Ref.~\cite{Shen:2025iue}. \label{fig8}}
\end{figure} 

The calculated binding energies and charge radii are presented in Figure~\ref{fig8}, alongside results from various nonrelativistic chiral potentials~\cite{Binder:2013xaa,Morris:2017vxi,Simonis:2017dny,Jiang:2020the,LENPIC:2022cyu} and experimental benchmarks~\cite{Wang:2021xhn,Angeli:2013epw}. The relativistic approach shows a remarkable capacity to reproduce both observables simultaneously. In contrast, nonrelativistic studies restricted to two-nucleon forces often struggle with a well-known tension: accurate binding energies lead to an underestimation of radii. Despite extensive efforts in nonrelativistic \textit{ab initio} modeling~\cite{Binder:2013xaa,Ekstrom:2015rta,Morris:2017vxi,Simonis:2017dny,Hoppe:2019uyw,Huther:2019ont,Ekstrom:2017koy,Hergert:2012nb,Heinz:2024juw,Elhatisari:2022zrb}, achieving a precise and concurrent description of these properties remains difficult, often due to significant uncertainties surrounding three-nucleon forces (3NFs). The relativistic framework effectively resolves this discrepancy without requiring explicit 3NFs, underscoring the vital role of relativistic effects in nuclear systems.

\begin{figure}[H]
%\isPreprints{\centering}{} % Only used for preprints
\includegraphics[width=8.0 cm]{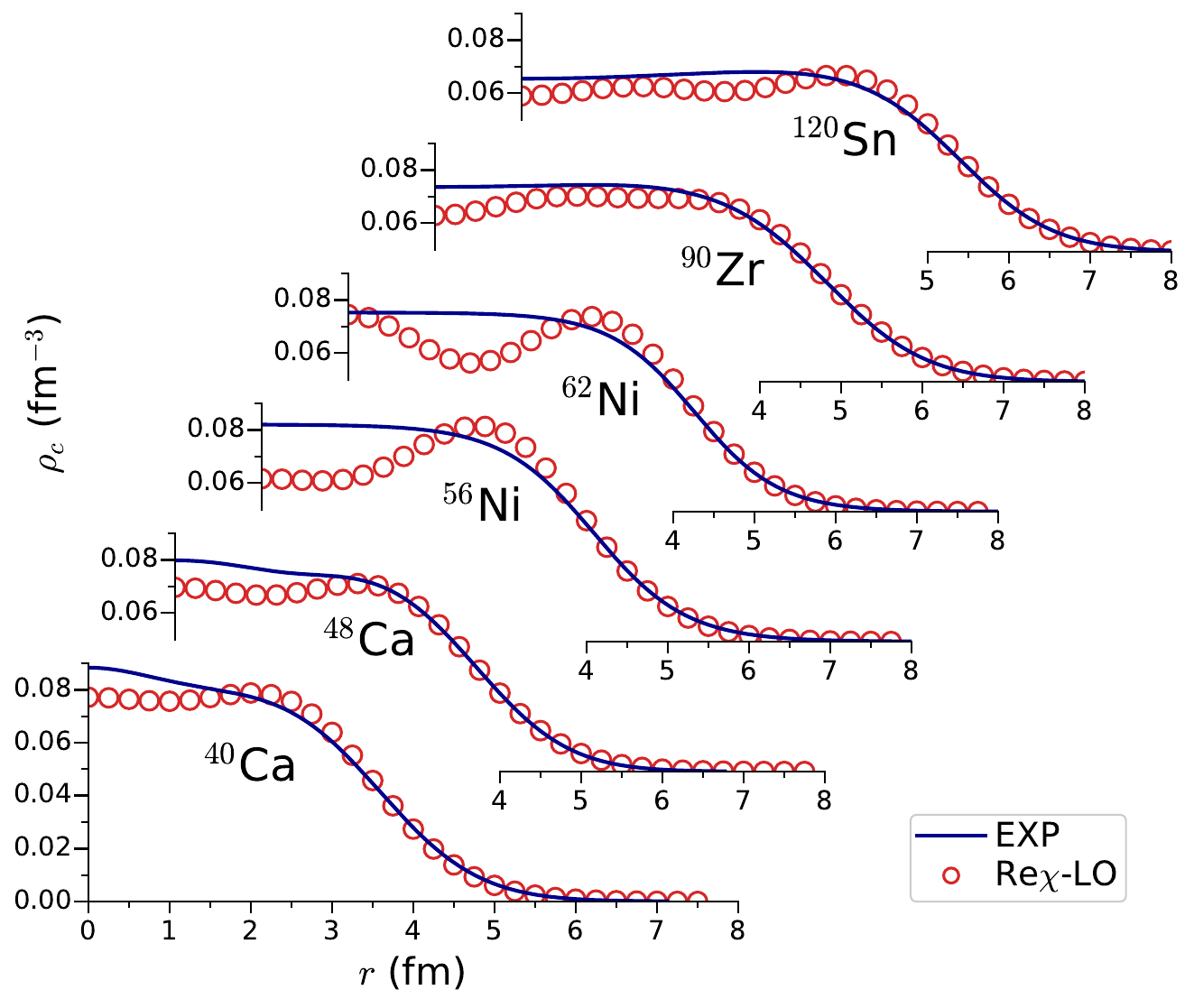}
\caption{Charge density distributions of nuclei ($A = 40$ to $120$) calculated using the RBHF theory with the LO relativistic chiral force, compared to experimental data~\cite{DeVries:1987atn}. Taken from Ref.~\cite{Shen:2025iue}. \label{fig9}}
\end{figure} 

Figure~\ref{fig9} illustrates the charge density distributions for nuclei with $A = 40$ to $120$, computed using the same LO relativistic chiral interaction featured in Figure~\ref{fig8}. The theoretical profiles show high fidelity to experimental data~\cite{DeVries:1987atn}. Notably, the central density for heavier isotopes saturates near empirical values, confirming that the saturation mechanism inherent in the relativistic framework is consistent across the nuclear chart. This establishes a coherent link between the saturation properties of finite nuclei and infinite nuclear matter~\cite{Zou:2023quo,Zou:2025dzh} within the context of this LO relativistic chiral approach.

\subsection{Hypernuclear systems}
In addition to nuclear systems, we have extended relativistic chiral nuclear forces to hyperon-nuclear (YN) forces and applied them to study hypernuclear systems. 
Our recent achievements in this area are:
\begin{enumerate}
\item An update of the relativistic chiral YN forces using physical baryon masses instead of average ones~\cite{Zheng:2025sol}, which is necessary for studying the in-medium interaction within the relativistic Brueckner-Hartree-Fock (RBHF) framework. The cross sections obtained with either physical masses or average masses are shown in Figure~\ref{fig:YNCS}. Our results for the $\Sigma^- p \to \Lambda n$, $\Sigma^- p \to \Sigma^0 n$, and $\Sigma^- p \to \Sigma^- p$ reactions show an overall downward shift compared to the case of average masses. The cross sections for $\Lambda p \to \Lambda p$ agree with the data even up to the $\Sigma N$ threshold. For $P_{\text{lab}}$ below 200 MeV/c, the agreement with the experimental central values is improved using physical masses compared to average masses. For the $\Sigma^- p \to \Lambda n$, the physical masses result in better agreement with the experimental data for $P_{\text{lab}}$ below 130 MeV/c.

\begin{figure}[H]
%\isPreprints{\centering}{} % Only used for preprints
\includegraphics[width=14.0 cm]{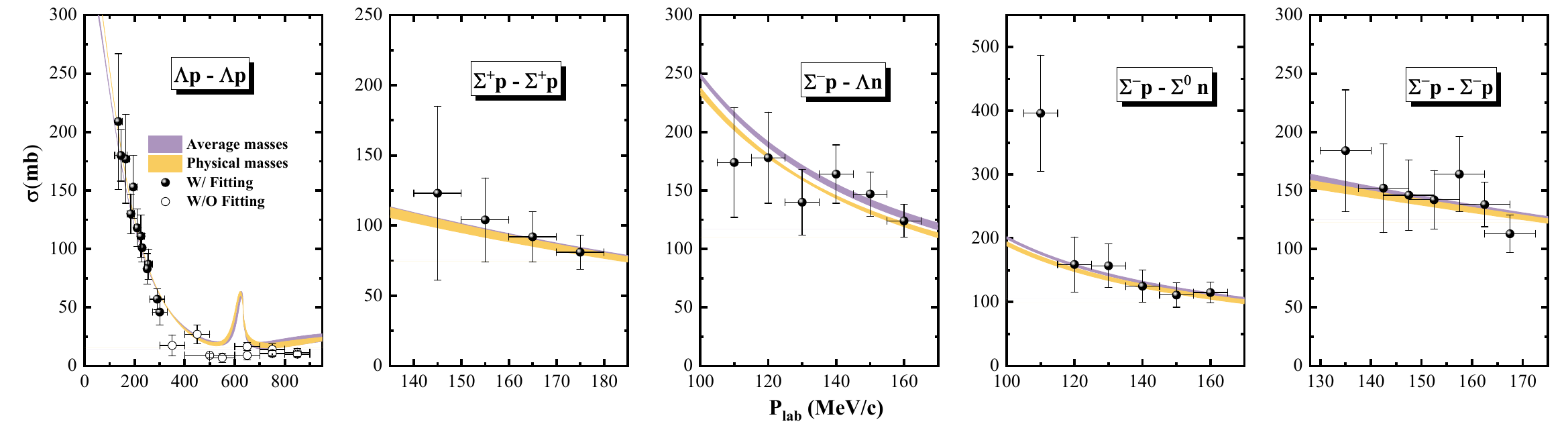}
\caption{Cross sections obtained with the LO covariant chiral YN interaction as functions of the laboratory momentum for $\Lambda=550-700$ MeV. The yellow bands represent the results obtained with physical baryon masses~\cite{Zheng:2025sol}, and the purple bands represent the results obtained with average baryon masses~\cite{Li:2016mln,Liu:2020uxi}. The experimental cross sections are from Refs.~\cite{Engelmann:1966npz,Sechi-Zorn:1968mao,Alexander:1968acu,Eisele:1971mk,Kadyk:1971tc,Hauptman:1977hr}. Taken from Ref.~\cite{Zheng:2025sol}.\label{fig:YNCS}}
\end{figure}

\item Calculation of the $\Lambda$ single-particle potential within the RBHF framework combined with the relativistic chiral YN interactions. As shown in Figure~\ref{fig:single}, the relativistic results are in better agreement with experimental data and other advanced theoretical models (such as Jul94) compared with the non-relativistic chiral YN force, where higher-order two-body chiral forces are generally required to achieve comparable agreement.

\begin{figure}[H]
%\isPreprints{\centering}{} % Only used for preprints
\includegraphics[width=8.0 cm]{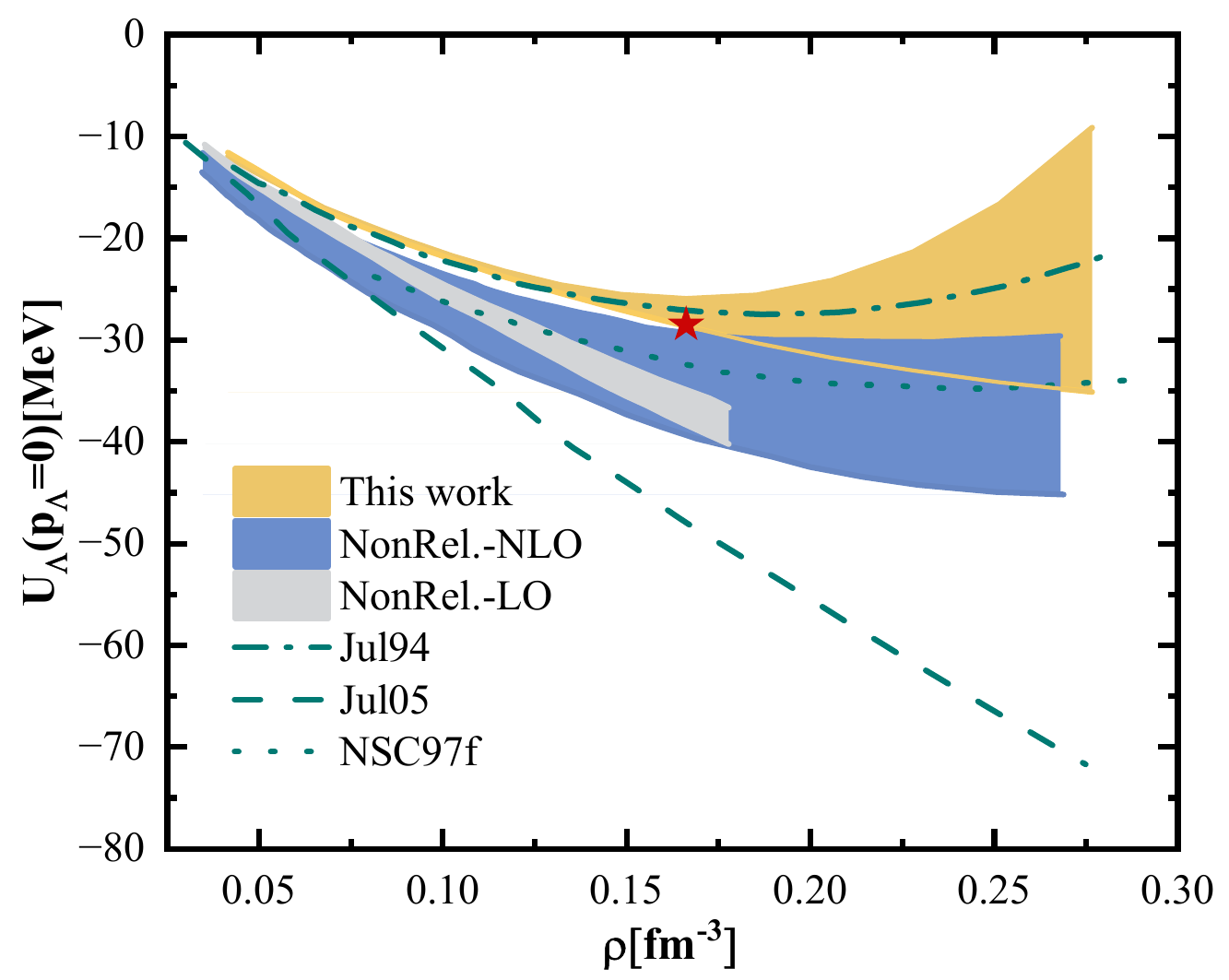}
\caption{The $\Lambda$ single-particle potential $U_\Lambda(p_\Lambda=0)$ as a function of density in SNM. The yellow band represents the results obtained with the LO covariant chiral YN interaction for cutoffs of $\Lambda_F=550$ (upper boundary) and 700 MeV (lower boundary), respectively. The blue band shows the NLO non-relativistic chiral YN interaction results for cutoffs of $\Lambda_F=500$ to 650 MeV, while the gray band is the LO non-relativistic chiral YN interaction results for $\Lambda_F=550$ to 700 MeV~\cite{Haidenbauer:2014uua,Haidenbauer:2019boi}. The dash-dotted curve is the result of the meson-exchange model Jul94, and the dashed curve is the result of the Jul05 potential~\cite{Hu:2014wja}. The dotted curve is the result of the Nijmegen NSC97f potential~\cite{Rijken:1998yy}, taken from Ref.~\cite{Yamamoto:2000jh}. The red star indicates the 'empirical value'. Taken from Ref.~\cite{Zheng:2025sol}.\label{fig:single}}
\end{figure} 

\item  Studies on hypernuclear structure within the Skyrme-Hartree-Fock approach combined with the above-mentioned in-medium interactions, including $\Lambda$ hyperon binding energies and energy levels of $\Lambda$ single-particle major shells, as shown in Figs.~\ref{fig:energylevel} and \ref{fig:binding}. The results indicate that, without adjustable parameters, relativistic chiral YN forces can provide a natural and accurate description of hypernuclear properties, which is crucial for understanding the structure and dynamics of hypernuclei.

\begin{figure}[H]
%\isPreprints{\centering}{} % Only used for preprints
\includegraphics[width=7.0 cm]{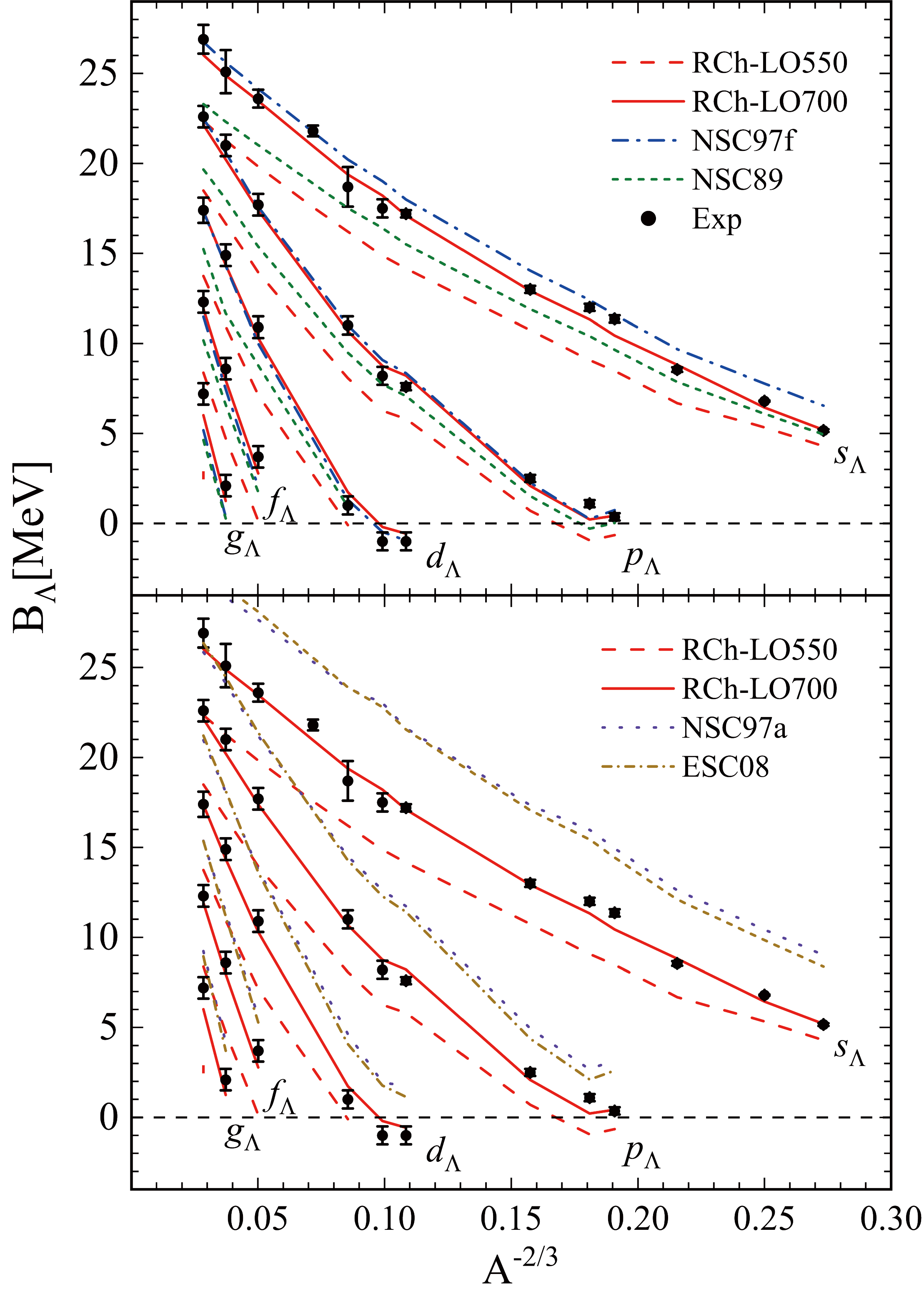}
\caption{Energy levels of the $\Lambda$ single-particle major shells in
$^A_\Lambda Z$ hypernuclei as a function of $A^{-2/3}$, calculated using the interactions RCh-LO550, RCh-LO700, NSC89, NSC97a, NSC97f, and ESC08, with experimental results included for comparison. Taken from Ref.~\cite{Zeng:2025fqb}. \label{fig:energylevel}}
\end{figure} 

\begin{figure}[htbp]
%\isPreprints{\centering}{} % Only used for preprints
\includegraphics[width=7.0 cm]{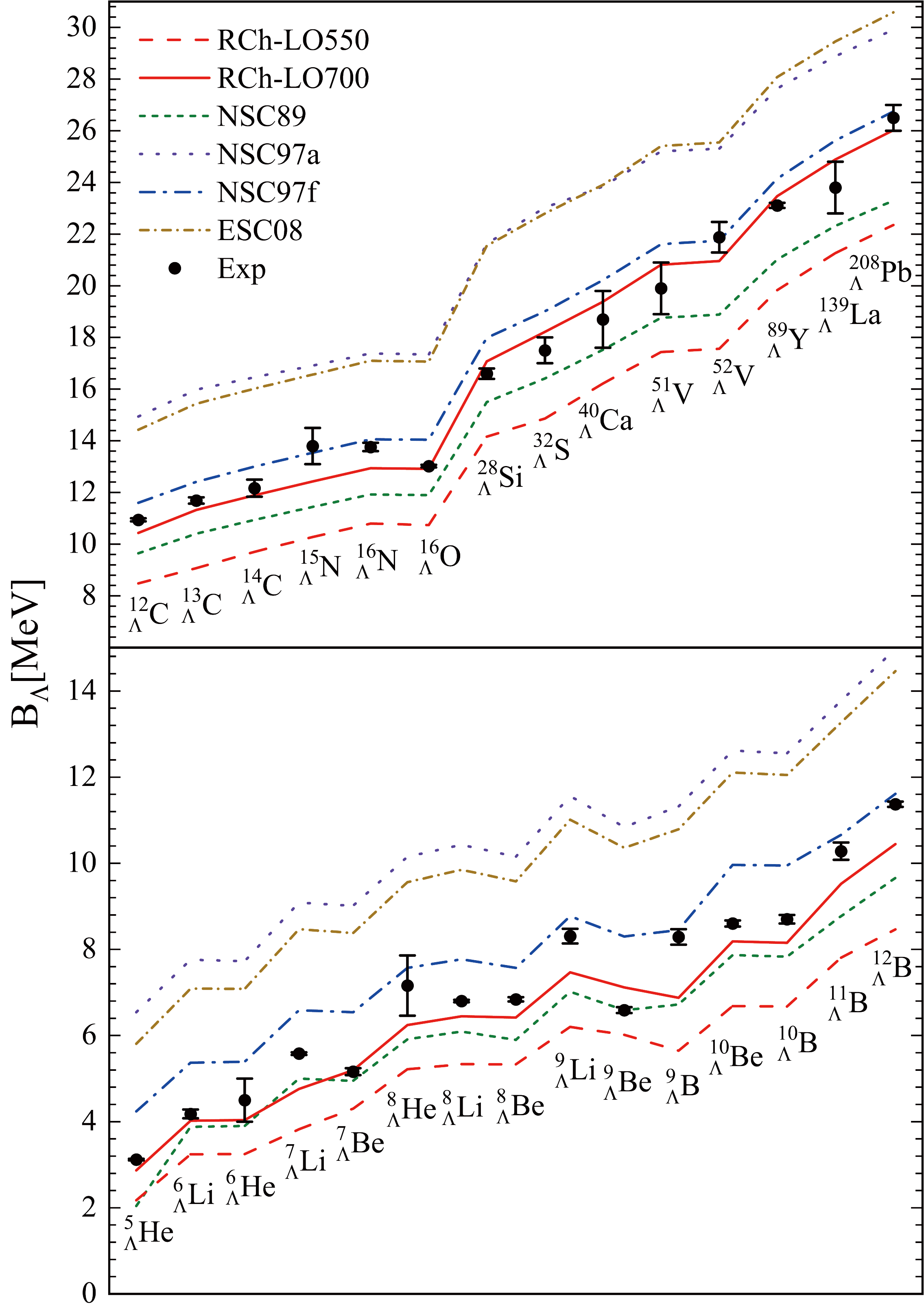}
\caption{$\Lambda$ hyperon binding energies across light to heavy mass regions calculated using the interactions RCh-LO550, RCh-LO700, NSC89, NSC97a, NSC97f, and ESC08, compared with experimental data. The red solid line represents the results obtained with our parameter set RCh-LO700, while the red dashed line corresponds to RCh-LO550. The green short-dashed line denotes NSC89, the purple dotted line indicates NSC97a, the blue dash-dotted line shows NSC97f, and the yellow short dot-dashed line represents ESC08. Taken from Ref.~\cite{Zeng:2025fqb}.\label{fig:binding}}
\end{figure} 
\end{enumerate}

\section{Summary and outlook}

In this review, we briefly summarize recent progress in relativistic chiral nuclear forces and their applications, integrating the latest developments through early 2026. The main achievements are summarized as follows:
\begin{enumerate}
\item We have clarified the necessity of relativistic/covariant theories for describing nuclear interactions, highlighting their advantages in satisfying Lorentz invariance and explaining fine structures in various systems. Relativistic frameworks have also been shown to offer a complementary approach to addressing longstanding challenges in \textit{ab initio} nuclear studies.

\item We have constructed the first high-precision relativistic chiral NN force up to NNLO, which exhibits improved renormalizability, faster convergence, and a more natural description than its non-relativistic counterparts.

\item We are currently constructing relativistic chiral NN forces up to N$^3$LO and have resolved the key technical complexities—including the handling of two-loop diagrams using the spectral representation method. 

\item We have applied relativistic chiral NN and YN forces to study symmetric nuclear matter, finite nuclei, and hypernuclear systems, yielding results that are in good agreement with experimental data. Our findings are consistent with recent relativistic \textit{ab initio} studies that have reproduced nuclear matter saturation and improved the "Coester line" for medium-mass nuclei.

\end{enumerate}

In the near future, we hope to address the following issues:
\begin{enumerate}
\item Complete the construction and fitting of relativistic chiral NN forces up to N$^3$LO, to check their convergence and improve the accuracy of high-energy scattering descriptions. 

\item Develop a fully self-consistent relativistic three-body scattering framework and apply it to understand $nd$ scattering and improve the description of three-body systems.

\item Extend relativistic chiral nuclear forces to isospin-asymmetric nuclear systems and rare isotope nuclei, and apply them to study the origin of heavy elements in astrophysics. We will further refine the treatment of isospin-breaking effects to more accurately describe charge-dependent nuclear properties.

\item Further explore the applications of relativistic chiral nuclear forces in hypernuclear systems and baryon-baryon scattering processes, and deepen our understanding of the strong interaction in low-energy regions. 
\end{enumerate}

\acknowledgments{The authors would like to thank all collaborators for their valuable contributions to this work. Special thanks go to the organizers and participants of Halo40 for valuable discussions on the latest progress in relativistic chiral nuclear forces. }

\funding{This research is partly supported by the National Natural Science Foundation of China under Grants No.12435007, No.11735003, and the Ministry of Education of the People's Republic of China. }

%%%%%%%%%%%%%%%%%%%%%%%%%%%%%%%%%%%%%%%%%%
\vspace{6pt}

\conflictsofinterest{
The authors declare no conflicts of interest.
} 
.

%%%%%%%%%%%%%%%%%%%%%%%%%%%%%%%%%%%%%%%%%%
%\isPreprints{}{% This command is only used for ``preprints''.
\begin{adjustwidth}{-\extralength}{0cm}

\reftitle{References}

%=====================================
\bibliography{refs}

\PublishersNote{}
%\isPreprints{}{% This command is only used for ``preprints''.
\end{adjustwidth}
\end{document}